\documentclass[useAMS,usenatbib]{mn2e}

\usepackage{graphicx,epsfig}  % got figures? uncomment this

\def\etal{\rm et al. \rm}

\title[Photometric calibration of the swift ultraviolet/optical
  telescope]{Photometric calibration of
  the {\it Swift} ultraviolet/optical telescope}

\author[T. S. Poole, A. A. Breeveld, and UVOT team members]{T. S.
Poole$^{1}$, A. A. Breeveld$^{1}$\thanks{E-mail: aab@mssl.ucl.ac.uk},
M. J. Page$^{1}$, 
W. Landsman$^{3}$, 
S. T. Holland$^{3,4,5}$, \and 
P. Roming$^{2}$, 
N. P. M. Kuin$^{1}$,    
P. J. Brown$^{2}$,
C. Gronwall$^{2}$, 
S. Hunsberger$^{2}$, \and 
S. Koch$^{2}$,
K. O. Mason$^{1,6}$,
P. Schady$^{1}$,  
D. Vanden Berk$^{2}$, 
A. J. Blustin$^{1}$, 
P. Boyd$^{3}$, \and
P. Broos$^{2}$,
M. Carter$^{1}$, 
M. M. Chester$^{2}$,  
A. Cucchiara$^{2}$,  
B. Hancock$^{1}$, \and
H. Huckle$^{1}$, 
S. Immler$^{3}$,
M. Ivanushkina$^{2}$, 
T. Kennedy$^{1}$,
F. Marshall$^{3}$, \and
A. Morgan$^{2}$,
S. B. Pandey$^{1}$, 
M. de Pasquale$^{1}$, 
P. J. Smith$^{1}$,  and M. Still$^{3}$ \\
$^{1}$Mullard Space Science Laboratory, University College London, Holmbury
  St. Mary, Dorking, Surrey, RH5 6NT, UK.\\
$^{2}$Department of Astronomy \& Astrophysics, Pennsylvania State
University, 525 Davey Laboratory, University Park, PA 16802, USA.\\
$^{3}$NASA/Goddard Space Flight Centre, Greenbelt, MD 20771, USA. \\
$^{4}$Universities Space Research Association. \\
$^{5}$Centre for Research and Exploration in Space Science and Technology. \\
$^{6}$Science \& technology Facilities Council, Polaris House, North Star Avenue, Swindon, Wilts SN2 1SZ, UK. }

\begin{document}

\date{Accepted ??? Received ????; in original form}

\pagerange{\pageref{firstpage}--\pageref{lastpage}} \pubyear{}

\maketitle

\label{firstpage}

\begin{abstract}

We present the photometric calibration of the {\em Swift} UltraViolet/Optical 
Telescope (UVOT) 
which includes: optimum photometric and background apertures, effective area
curves, colour transformations, conversion factors for count rates to flux, and the 
photometric zero points (which are accurate to better than 4 per cent) for
each of the seven UVOT 
broadband filters. The calibration was performed with observations of standard
stars and standard star fields that represent a wide range of spectral star
types. The calibration results include the position 
dependent uniformity, and instrument response over the 1600--8000\AA~  
operational range. Because the UVOT is a photon counting instrument, we also 
discuss the effect of coincidence loss on the calibration results. We
provide practical guidelines for using the calibration in UVOT data
analysis. The results presented here supersede previous calibration results.

\end{abstract}


\section{Introduction}
\label{Intro}
The Ultraviolet Optical Telescope \citep[UVOT;][]{RPWA2005} is a modified
Ritchey-Chr\'{e}tien telescope with a $17 \times 17$ arcmin field-of-view
(FOV) operating in the 1600--8000\AA~range. Like many optical
telescopes, the UVOT uses broadband filters in the ultraviolet (UV) and optical to obtain
colour information. However, unlike most optical telescopes, the UVOT has a
photon counting detector that gathers data in a similar way to an X-ray detector.

The UVOT is one of three telescopes flying on board the {\it Swift}
spacecraft \citep{GN2004} and is co-aligned with the $15-150 {\rm ~keV}$ Burst
Alert Telescope \citep[BAT;][]{BS2005} and the $0.2-10 {\rm ~keV}$ X-Ray
Telescope \citep[XRT;][]{BDN2005}. The primary goal of the {\it Swift}
mission is to detect and characterize gamma-ray bursts (GRBs) and their
afterglows. The design of the UVOT is well suited to this goal.
In addition, UVOT's UV response in particular makes it a valuable instrument for other
types of observations.

There are currently four near-UV imaging telescopes operating in space: {\it HST}
- Wide-Field Planetary Camera 2 \citep[WFPC2;][]{BCJ1994}, {\it XMM}-Optical
Monitor \citep[OM;][]{MK2001}, {\it Galaxy Evolution Explorer} \citep[GALEX;]
[]{SOHW2004,MB2001,BL2000}, and {\it Swift} - UVOT. The HST has the highest
spatial resolution and sensitivity but the smallest FOV. GALEX has the
largest FOV and covers a larger UV wavelength range than OM or UVOT but it
has very broadband filters and the lowest spatial resolution of the four
telescopes. The UVOT and OM are very similar in design, both covering
approximately the same wavelength range and containing similar broadband
filters in both the UV and optical regimes. Both UVOT and OM include
three UV filters that together cover approximately the same band-pass as the long wavelength GALEX filter. The UVOT has a slightly broader point spread function than OM 
but a higher
sensitivity, by a factor of $\sim10$ at the bluest wavelengths. With respect
to the four telescopes, UVOT, being mounted on {\em Swift}, is able to 
respond most quickly, making it the
best instrument for observing transient phenomena.

In this paper, we describe the in-orbit photometric calibration of the UVOT and
define the UVOT photometric system.
We begin by describing the UVOT technical details in Section~\ref{tech}, the 
calibration strategy in Section~\ref{strategy}, followed
by the observational measurements and photometry method in Sections~\ref{gb
  meas}--\ref{photometry}. 
We review the calibration of the coincidence loss and 
position-dependent uniformity in Sections~\ref{coi} and~\ref{uniform},
respectively, and we provide the in-orbit effective area calculations, photometric zero points,
colour transformations, and flux conversion factors in Sections~\ref{eff
  area}--\ref{flux conv}. We test the calibration in Section~\ref{verification} and give guidelines for making use of the calibration for UVOT data analysis in Section~\ref{practical}. Finally we summarise the calibration and 
discuss what is left to be done in Section~\ref{discussion}. 

Our preliminary UVOT calibration, described in \citet{b5} and \citet{b6}, was released soon
after launch in the HEASARC calibration database
(version 20050805 of the Swift/UVOTA 
CalDB\footnote{All the CalDB files can be found at \newline 
http://swift.gsfc.nasa.gov/docs/heasarc/caldb/swift/}). 
\citet{LW2006} later gave an 
independent calibration for the optical
filters using data from early in the mission. The
calibration analysis described in this paper is the most detailed and 
comprehensive to date and thus supersedes our previous work. 

\section{UVOT Technical Details}
\label{tech}
A technical description of the UVOT is given by \citet{RPWA2005}; we summarise
the relevant properties here. 
The UVOT is of a modified Ritchey-Chr\'{e}tien design, with a 30cm primary 
mirror and an {\it f}-ratio of $f/12.7$ after the secondary. A 45 degree 
mirror directs photons into one of two detector units, one of which is kept
in cold redundancy. Each detector unit consists of a filter wheel and 
micro-channel plate intensified CCD \citep [MIC;][]{k1,f1}. The filter wheels contain 11 
slots which house 3 optical filters, 3 UV filters, a clear
{\it white light} filter (which transmits throughout the UVOT wavelength 
range of 1600--8000\AA), a magnifier, a low-resolution optical grism, a low-resolution UV
grism, and a blocked filter. The characteristics of the optical and UV 
filters can be found in Table~\ref{tabPete}. 

The MIC detector has an entrance window that is slightly 
figured optically to flatten the image plane of the telescope. An S20 photocathode is deposited on the inside of the window and is optimized for the UV and blue wavelengths. The photocathode converts an incoming photon into an electron signal, which is then amplified by a factor of a million by a photomultiplier stage. Thus every incoming photon results in a cloud of electrons at the back of the photomultiplier, and these are converted back into photons by a phosphor screen. The resulting photon splash is recorded on a fast scan CCD, which has an active area of $256 \times 256$ pixels. Each photon splash extends over several CCD pixels, and this allows the centroid of the splash to be determined in real time by fast onboard electronics to a fraction of a CCD pixel  
\citep[$1/8^{th}$ of a CCD pixel in the case of the UVOT MIC detectors; ][]{MR1997}. The resulting image format is thus $2048 \times 2048$ pixels, covering a field of view of $17 \times 17$ arcmin. This provides a spatial sampling of 0.5 arcsec. There is a fibre taper between the phosphor screen and the CCD, which compensates for the larger physical area of the photomultiplier stage compared to the CCD. Throughout this paper, the word {\it pixel} refers to $0.5 \times 0.5$ arcsec image pixels, unless explicitly stated otherwise.

\begin{table}
\caption{{\it Swift}/UVOT Filter Characteristics. The central wavelength is the
midpoint between the wavelengths at half maximum}
\label{tabPete}
\begin{tabular}{@{}lcc}
\hline
Filter & Central Wavelength & FWHM \\
 & (\AA) & (\AA) \\
\hline
v & 5468 & 769 \\
b & 4392 & 975 \\
u & 3465 & 785 \\
uvw1 & 2600 & 693 \\
uvm2 & 2246 & 498 \\
uvw2 & 1928 & 657 \\
\hline
\end{tabular}
\end{table}

The advantages of the MIC detector over a traditional `bare' CCD are that (a) it is photon counting, (b) it operates at `room temperature' without the need for cooling and (c) it is insensitive to cosmic ray and charged particle hits in the CCD. Because it is a photon counting detector, the UVOT suffers from
{\it coincidence losses} at high photon rates, when 
two or more photons arrive at a similar location on the detector within the
same CCD read out interval \citep {f2}. It is equivalent to `pile-up' in an
X-ray CCD detector. The magnitude of the effect depends on the read-out rate of
the CCD, which is once every 11.0329 ms when the full $256 \times 256$ CCD pixels are
used, and faster when only part of the field is read out. In addition to
coincidence loss the count rates also have to be corrected for the
{\it dead time} while the charge is transferred out of the CCD, which amounts 
to 1.6 per cent of the full-frame read-out interval. The loss due to dead time
is a constant factor for a given read-out rate,
whereas coincidence loss also depends on the incoming photon rate.

For the full CCD, with a read-out time of $\sim 11 {\rm ~ms}$, coincidence losses start to be significant at $\sim 10 {\rm ~counts\,s^{-1}}$ and a correction should be applied to the recorded signal. Beyond $\sim 90 
{\rm ~counts\,s^{-1}}$, equivalent to 1 count per frame, correcting for
coincidence losses becomes increasingly uncertain. To extend the range of the
detector to brighter sources, the read-out rate can be increased, as noted
above, by using only a subset of the CCD pixels. We use two pre-defined
hardware windows, which are $120 \times 120$ CCD pixels ($8 \times 8$ arcmin)
and $75 \times 75$ CCD pixels ($5 \times 5$ arcmin) on a side, centred on the
observatory boresight. These reduce the frame times to 5.417ms and 3.600ms,
respectively, and also reduce the dead time by a small amount.

The onboard algorithm that centroids the photon splash recorded on the MIC CCD
uses a simple look-up table for speed. The intrinsic imperfections in this
process mean that the sub-pixels within each physical CCD pixel do not all have
the same effective physical area. The signature of this in the resulting image
is an apparent $8 \times 8$ pixel fixed-pattern effect,
sometimes referred to
as `mod-8 noise', though it is not strictly noise since photons are 
conserved \citep{MR1997}. A calibration of this is derived onboard based 
on illuminating
the field with an LED within the instrument, and applied as a correction to the
centroiding algorithm. Nevertheless small gain variations over the face of the
detector mean that a low-level residual fixed-pattern effect is still seen
which varies in magnitude depending on position on the detector. If desired, the effect can
be largely removed on the ground by using the tool {\tt uvotmodmap}, which is
released as part of the {\it HEAsoft Swift} ftools software
package\footnote{HEAsoft software can be found at \newline
http://heasarc.gsfc.nasa.gov/docs/software/lheasoft/}.

The UVOT operates in two data-taking modes: event and image. Event mode preserves the
temporal and positional information of each photon. In image mode the photons
are accumulated into an image in the instrument's on-board memory before being
telemetered to the ground, thus providing positional
but not temporal information on the photons. 

Because the spacecraft can drift slightly during an observation, the photon positions are shifted before they are added to memory. Occasionally, due to large spurious drifts in the 
spacecraft attitude information, individual frames are not added to the image. This time lost, when the 
onboard shift-and-add algorithm unnecessarily tosses events off 
the image, is known as TOSSLOSS. TOSSLOSS occurred fairly often until a software fix
in September 2005, but scarcely ever since then. 

\section{Calibration Strategy}
\label{strategy}
In the following sections we describe the different elements of the UVOT 
calibration in detail, but first we will outline the overall approach that we
have taken.
As a starting point, we used our ground-based measurements 
of the various components in the UVOT optical path, to produce an idealised or 
reference effective area curve for each of the UVOT filters. From these
effective area curves, we predicted the count rates for a number of photometric
standard stars for each of the UVOT filters. We then compared these
predictions with in-orbit measurements of the standard stars, to produce an
in-orbit correction curve to the overall instrument response as a function of
wavelength. Our predicted effective area curves were multiplied by the
in-orbit correction curve to obtain the in-orbit effective area curves.
Once the in-orbit effective areas were established, zero
points, colour transformations, flux conversion factors etc.\thinspace
followed. 

Any changes in the overall filter transmission are thus incorporated into 
the correction curve. Changes in filter transmission shape would be more difficult 
to deal with, but would show up as a difference in zero point for sources of 
different colour. We do not find evidence for a change in the transmission 
shape for any of the filters.

\section{UVOT Instrument Response}
\label{gb meas}

The instrument response of the UVOT is a product of:
\begin{enumerate}
\item the telescope primary mirror geometric collecting 
area of 659cm$^2$, 
\item the mirror reflectivity \citep{b16},
\item the filter transmission curves,
\item the detector quantum efficiency (DQE), which is the overall sensitivity of the photon counting system, including the photocathode sensitivity~\citep{b15}. 
\end{enumerate}

The DQE and filter transmission curves (for all but the white filter, see
below) as a function of wavelength were measured in the laboratory at a
sub-unit level. The mirror reflectivity was also determined in the laboratory,
by measuring planar witness samples that were coated alongside the mirrors. 

\begin{figure}
\includegraphics[angle=90,width=84mm]{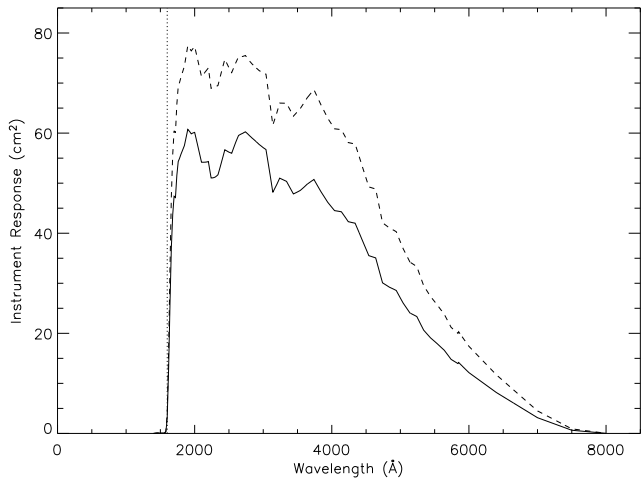}
\caption{The instrument response of the UVOT including detector window, cathode
  sensitivity, mirror reflectivity and telescope area. The dashed line
  represents the idealised instrument response (assuming all photons are collected in the image) while the solid line shows the actual 
  in-orbit response measured in a 5 arcsec radius circular aperture. The vertical dotted line marks the short wavelength cut-off of the instrument at 1600\AA.}
\label{figInstResp}
\end{figure}

The measurements listed above, but not including the filter transmission curves,
were combined to produce an idealised, or `reference' response for the
telescope, which is shown as a dashed line in Figure~\ref{figInstResp}. 
This is effectively the response expected if all the photons that pass 
through the telescope can be captured in the image. In practice, residual 
imperfections in the reflecting surfaces will scatter some photons into 
broad wings in the telescope point spread function, and thus the actual 
throughput for photons in the image core will be less than ideal.

The dotted line in Figure~\ref{figInstResp} is placed at the wavelength of 1600\AA, which is the short wavelength limit of the instrument response determined by the detector window transmission. The long wavelength cut-off has been set at 8000\AA. Outside the limits of 1600\AA~and 8000\AA~the
detector quantum efficiency is very small; this range includes 99.98
per cent of the total instrument response.

\begin{figure}
\includegraphics[angle=90,width=84mm]{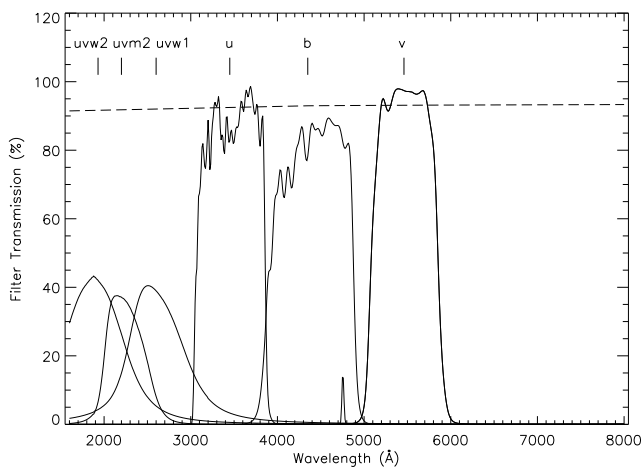}
\caption{Filter transmission curves for the UVOT filters as measured in the laboratory. The white filter
  transmission curve is given by the dashed line; the identities of the other filters are
  indicated on the plot. The extra peak at $\sim4750$\AA~is part of the u filter transmission.}
\label{FigFT}
\end{figure}

Figure~\ref{FigFT} shows the transmission curves of the optical and UV filters as 
measured in the laboratory \citep{b18}. The extra peak at
$\sim4750$\AA~is part of the u filter transmission. The error at each point 
on these transmission curves is at most $\pm 1$ per cent (accounting for
systematics), but is probably $\pm 0.4$ per cent. The
curves provided for uvw2 do not extend shortward of 1800\AA~due to the
inability of the measuring device to provide data below this
wavelength; therefore this filter curve had to be extrapolated to the
1600\AA~limit assuming a peak at 1930\AA, and a symmetrical response profile.
There were no laboratory measurements of the white light filter 
transmission as a function of wavelength. Instead, the curve that we have
adopted, and which is shown in Figure~\ref{FigFT}, is a model of the 
transmission, based on the design and substrate properties of the white filter. 

The predicted effective area curves for each of the filters were obtained
by combining the instrument response curve (Figure~\ref{figInstResp}) with the 
individual filter transmission curves (Figure~\ref{FigFT}).

\section{Standard star data}
\label{standards}

Table~\ref{tabSources} lists the standard stars and standard 
fields that were used for the calibration work presented in
this paper. In total, two Oke standard stars, six white dwarf standards,
ten Landolt standard stars, and four star fields have been used for the various
calibration functions listed in Table~\ref{tabSources}.

For the standard stars used to determine the in-orbit 
effective areas, zero points, colour transformations
and count rate to flux conversion factors, our calibration procedure requires 
spectra that can be folded through the instrument response and filter
transmission curves. It has been a challenge to find spectrophotometric standard
stars that are known to sufficient accuracy, have wide enough spectral
coverage, are not variable, and are not too bright for the
sensitive UVOT detectors. Thus we have been limited in the UV to just 3 trusted
white dwarf standards. For the optical we have used photometric standards to
increase the sample of sources.

\begin{figure}
\includegraphics[angle=90,width=84mm]{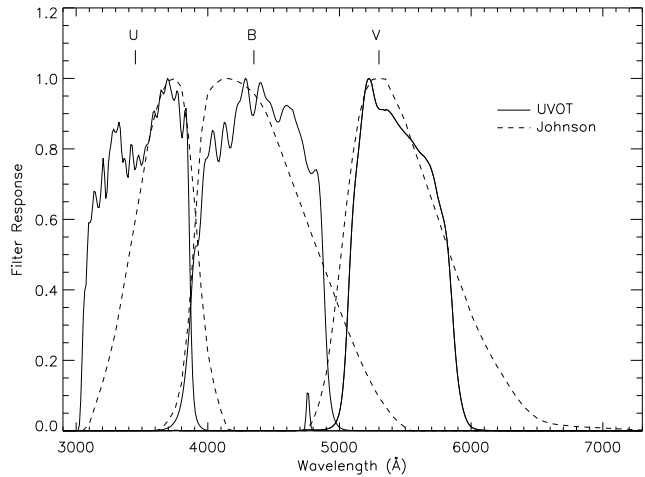}
\caption{Comparison between the Johnson and UVOT optical responses. The Johnson responses are taken from \citet{b8} and \citet{b9}, as described in Section~\ref{standards}.}
\label{FigFILcomp}
\end{figure}

Representations of the Johnson UBV passbands are required for several 
aspects or our calibration (e.g. normalisation of standard star spectra as
described below and colour transformations from UVOT to the Johnson
system). However, the true shapes of the Johnson UBV response curves 
have long been
controversial \citep{bessel05}. In this work, we have taken the following
response curves to represent the Johnson UBV system. For the V and B responses 
we have
used Table 1, columns ``$\phi_V$'' and ``$\phi_B$'' respectively, 
from \citet{b8}. For the Johnson U response, we
have used Table 2 from \citet{b9}. Figure~\ref{FigFILcomp} shows the comparison
between the Johnson U, B and V and UVOT
normalised u, b and v responses. In this paper we use the convention of capital letters for the Johnson system
magnitudes and lower case letters for the UVOT system magnitudes.

For the two Oke standard stars, spectra were obtained from the ESO standard
star archive (ftp://ftp.eso.org/pub/stecf/standards/hststan). These spectra
cover the wavelength range 3200--8000\AA~and are therefore only used to
calibrate the v and b
filters. As described in \citet{b10}, the absolute photometric calibration of
these spectra can be improved using precision ground-based photometry. 
These spectra were therefore folded through the Johnson V
filter response and normalised to the V magnitudes given by \citet{b10}.

Three of the white dwarf standards (WD1657+343, 
WD1026+453, WD1121+145) are used for the effective areas and zero point
calibration. 
Spectra were obtained from the HST MAST archive
(http://archive.stsci.edu/) of WD1657+343\footnote{MAST IDs: 08v101010,
  08h111010, 08h111040, 08v101030} and WD1026+453\footnote{MAST IDs: 08h106040,
  08h106010, 08h106020}. The spectrum of WD1026+453 had to be extrapolated
longward of 5700\AA, and so this spectrum was not used for calibration of the v
filter. For WD1121+145 we use the IUE spectrum
presented by \citet{holberg03} and the optical spectrum from \citet{massey88}.
For these three sources, the Johnson magnitudes listed in Table~\ref{tabSources} 
were obtained by folding their spectra through the Johnson UBV response curves.

The Landolt standard stars \citep{b1} do not have accurate spectra available in the
literature. Therefore, the Landolt standard stars were matched to the
stellar spectra from the catalogue of \citet{b2} which had the closest 
B-V and U-B colours. The best-matching Pickles spectra were
then folded through the Johnson transmission curves and normalised to the
U, B, or V magnitudes from \citet{b1} for calibration of the UVOT u, b and v filters 
respectively.

\subsection{In-orbit measurements}
\label{In-Orbit Meas}

Table~\ref{tabSources} gives observational 
details for all of the calibration sources and fields, including observation 
dates (Column 2), 
the observed filters (Column 3), and the photometric calibration product in which 
the source or field was used (Column 4). Most of the observations were taken 
between launch and April 2005,
during the calibration phase of the {\it Swift} mission. For more details
about the observational information please refer to the  {\it Swift} Calibration
Database (CalDB) documents\footnote{
http://swift.gsfc.nasa.gov/docs/heasarc/caldb/swift/}~\citep{b22,b23,b24}. 
The original ground-based UBV photometry for the Landolt standard stars 
was obtained using a 14 arcsec diameter aperture \citep{b1}, which is not 
optimum for the UVOT calibration. We therefore checked each of the UVOT images of 
Landolt standard stars to ensure that there were no other stars of sufficient 
brightness to offset the photometric calibration, 
within 7 arcsec of the standard stars.  

\begin{table*}
\caption{Observational data used for the UVOT photometric calibration. The 
uses are given in column 4 where: OA is
optimum photometric aperture (Section~\ref{opt ap}), BA is background aperture (Section~\ref{back meth} \&~\ref{back size}), CL is coincidence loss empirical correction 
(Section~\ref{coi}), PDU is
  position dependent uniformity (Section~\ref{uniform}), EA is in-orbit effective area
  curve (Section~\ref{eff area}), ZP is photometric zero points (Section~\ref{zp}), and CT is
  colour transformations (Section~\ref{colour trans}). The last three columns show the Johnson
  V, B and U magnitudes: for the Landolt sources from \citet{b1}, the Oke from \citet{b10} and
  white dwarfs calculated as described in Section~\ref{standards}.}
\label{tabSources}
\begin{tabular}{@{}llllllll}
\hline
Source & Date & Filter & Use & Origin & V & B & U \\
\hline
SA95-42 & July 05 & v & EA, ZP, CT & Oke & 15.606	& 15.391 & - \\
g24-9 & July 05 & v, b & EA, ZP, CF & Oke &15.751	& 16.176 & - \\
WD1657+343 & Feb-April 05, June 05 & v, b, u, uvw1, uvm2, uvw2, white & EA, ZP & White Dwarf
& 16.4	& 16.2 &   15.0
\\
WD1026+453 & July 05, Oct-Nov 05 & b, u, uvw1, uvm2, uvw2 & EA, ZP & White Dwarf & 16.1 	& 15.9 &   14.8
\\
WD1121+145 & Feb-May 05 & uvw1,  uvm2, uvw2, white & BA, EA, ZP & White Dwarf & 16.9	& 16.6 &   15.4
\\
GD128 & Nov-Dec 05, Jan 06 & white & OA, BA & White Dwarf & 15.89	& 15.82 & -
\\
Hz2 & Nov-Dec 05 & v, b, u, uvw1, uvm2, uvw2 & OA, BA & White Dwarf & 13.86	& 13.81	 & -
\\
GD50 & Dec 05-June 06 & v & PDU & White Dwarf  & 13.98	& 13.82	 & -
\\
SA95-102 & March 05 & v, b, u & EA, ZP, CT & Landolt & 15.622  & 16.623 & 16.785
\\
SA98-646 & March 05 & v, b, u & EA, ZP & Landolt & 15.839  & 16.899 & 18.325
\\
SA101-278 & March 05,  Nov 05 & v, b, u, white & EA, ZP, CT & Landolt & 15.494  & 16.535 & 17.272
\\
SA101-13 & March 05, Nov 05 & v, b, u, white &  EA, ZP, CT & Landolt & 15.953  & 16.590 & 16.557
\\
SA104-244 & Feb-March 05 & v, b, u, white & EA, ZP, CT & Landolt & 16.011  & 16.601 & 16.449
\\
SA104-338 & Feb-March 05 & v, b, u, white & EA, ZP, CT & Landolt & 16.059  & 16.650 & 16.568
\\
SA104-367 & March-April 05 & v, b, u & EA, ZP, CT & Landolt & 15.844  & 16.483 & 16.357
\\
SA104-443 & March-April 05 & v, b, u & EA, ZP, CT & Landolt & 15.372  & 16.703 & 17.983
\\
SA104-457 & March-April 05 & v, b, u & EA, ZP, CT & Landolt & 16.048  & 16.801 & 17.323
\\
PG1525-071b & March 05 & v, b, u & EA, ZP & Landolt & 16.403  & 17.133 & 17.268
\\
NGC 188 & Oct 05 & b & CL & Open cluster & -	& -	 & -\\
SA95 & July-October 05 & b & CL & Star field & - & - & -\\
M67 & March 06 & b & CL & Open cluster & - & - & -\\
SA104 & March 05 & b & CL & Star field & - & - & -\\
\hline
\end{tabular}
\end{table*}

\section{Photometry Method}
\label{photometry}

In this section we describe how and why we chose the photometric aperture and 
background region to use for the calibration analysis, and describe the 
analysis we undertook. We do not necessarily recommend following the same 
procedure for all UVOT data analysis and the reader is referred to
Section~\ref{practical} for more information on this.

Much of this analysis was performed using in-house IDL\footnote{from ITT Visual
Information Solutions} routines which could be 
readily tailored to the calibration tasks at hand, and have allowed us to test 
analysis methods and calibration products that have subsequently been implemented in 
the official UVOT HEAsoft software and HEASARC calibration database.

\subsection{Optimum photometric aperture}
\label{opt ap}
One effect of coincidence loss (see Section~\ref{coi}) is that the shape of the point spread function
(PSF) is slightly dependent on the source count rate. We wished to find an
aperture enclosing the same percentage of the PSF regardless of count rate, 
if such an aperture existed. This was most easily found by using
the same source and changing the frame rate to mimic a change in count 
rate and then looking for a consistent result after correcting for coincidence
loss (see Section~\ref{coi}, Equation~\ref{Equ3}). 
The use of different hardware window sizes, and hence frame times, 
allows coincidence loss to be varied without changing any other parameters. 
Therefore, we define the optimum photometric aperture radius to be 
that which gives rise to the smallest variation in the enclosed energy 
fraction for a point source as a function of coincidence loss. 
This is the most convenient aperture to use for the calibration
because it gives consistent results for photometry over a wide range of count rates.

\begin{figure}
\includegraphics[angle=0,width=84mm]{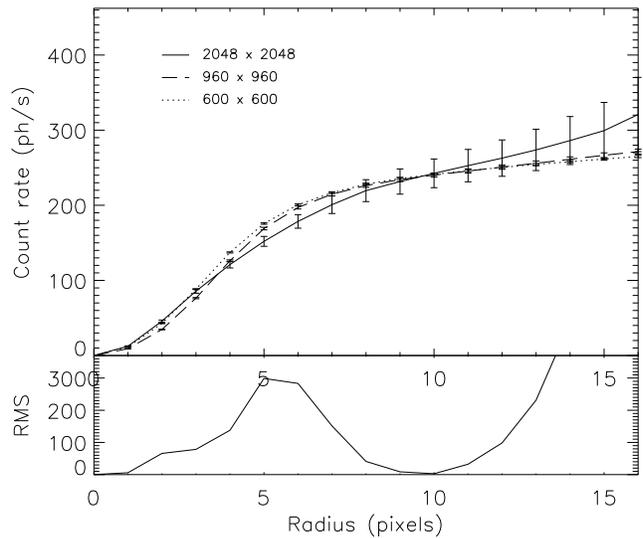}
\caption{Coincidence-loss-corrected count rates within concentric radii, for Hz2 in the uvw2 filter, with a hardware window size of $2480 \times
  2480$ (solid black line), $960 \times 960$ (dashed line), and $600 \times
  600$ (dotted line). The error bars on the background subtracted data were
  determined using Poisson statistics}
\label{FigE}
\end{figure}

The optimum aperture was investigated using observations of two isolated,
bright stars GD128 and Hz2, for the default set of hardware windows with sides of length
2048, 960 and 600 pixels. Hz2 was used for the optical and UV filters, but is
too bright for the white filter, for which GD128 was used.
The raw count rates were corrected using the theoretical coincidence loss
equation only (see Section~\ref{coi}, Equation~\ref{Equ3}), which takes the
frame time into consideration. Figure~\ref{FigE} shows an example of this for the 
uvw2 filter. The minimum of the RMS curve coincides with the optimum aperture radius.

Table~\ref{tabOptAp} shows the results from Hz2 (optical and UV filters) and GD128 (white filter) for the optimum aperture: the
optical, UV and white filters have optimum aperture radii of $10.1\pm 1.3$ pixels,
$10.7\pm 1.5$ pixels, and $9.8\pm 1.3$ pixels respectively. Since these are 
similar, and consistent within the uncertainties with a mean of $10.5\pm 1.2$ 
pixels, we chose to use a radius of 10 pixels (5 arcsec) for all the 
photometric calibration analysis.

\begin{table}
\caption{Optimum aperture results obtained from Hz2 (optical and UV filters) and GD128 (white filter).}
\label{tabOptAp}
\begin{tabular}{@{}lcc}
\hline
Filter & Aperture radius (pixels) & Range (pixels) \\
\hline
v & 11.7 & 2.6 \\
b & 10.8 & 1.2 \\
u & 8.7 & 1.8 \\
uvw1 & 12.5 & 0.9 \\
uvm2 & 10.0 & 0.5 \\
uvw2 & 9.7 & 0.3 \\
white &  9.8 & 1.3 \\   
\hline
\end{tabular}
\end{table}

The percentage of counts within the chosen 10 pixel aperture radius is
stable for moderate count rates. Assuming that a 55 pixel radius
aperture represents 100 per cent of the PSF (see Section~\ref{back size} 
and \citet{bALICE}), then $85.8
\pm 3.8$ per cent of the PSF is contained within the 10 pixel radius for all filters 
including the white filter. The differences between the enclosed energy fraction within 
our adopted 10 pixel radius aperture, and the optimum apertures we determined for 
the individual filters (Table~\ref{tabOptAp}) are very small
because this radius is already in the wings of the PSF where the count
rates are low (Figure~\ref{FigEXT}). In the case of uvw1, which has the largest 
difference between the optimum aperture radius of 12.5 pixels, 
and our adopted aperture radius of 10 pixels, the difference in the enclosed energy 
fraction is at most 2 per cent.

\subsection{Subtracting the background}
\label{back meth}
Two methods of determining the background count rate were considered: mean background, or clipped-mean background. The mean background
method averages the number of counts per pixel over the background
aperture, whereas the clipped-mean background method excludes any background pixels with
count rates more than
3 sigma above the initial mean and then averages the counts over the remaining
background aperture pixels.  
The advantage of the
clipped-mean background method is that it removes counts from any significant sources that may lie within the background aperture. However, the disadvantage is that it produces a consistently lower value because it removes the top of the
distribution, which is assumed to be Gaussian. For very low rate
backgrounds, as is often the case with UVOT, this assumption is not valid: most
pixels contain either 1 or 0 counts.

Using the software package GAIA (starlink GAIA version 2.8-0) the clipped-mean
method was consistently smaller than the mean method by $0.14\pm 0.03$
ph/pix. An in-house IDL routine was also used, and also resulted in a
consistently lower background value when using the clipped-mean method.
This is acceptable when observations have high backgrounds, but will have a
significant effect on count rates for observations with lower backgrounds and
faint sources. Hence, a background limit of
10 ph/pix was set (where the measured background difference is 1.4 per cent of the total
background), above which the clipped-mean background method is applied, and below
which the simple mean background method is used unless there is an obvious 
source that lies within the background region.

\subsection{Background region size}
\label{back size}
For bright sources PSF wings can be seen that extend out
to around $40-55$ pixels. Investigation into the extent of the PSF wings was
carried out by examining images of three
white dwarf standard stars (Hz2, GD128 and WD1121+145) of different 
magnitudes, with the three default hardware window sizes. 

\begin{figure}
\includegraphics[angle=90,width=84mm]{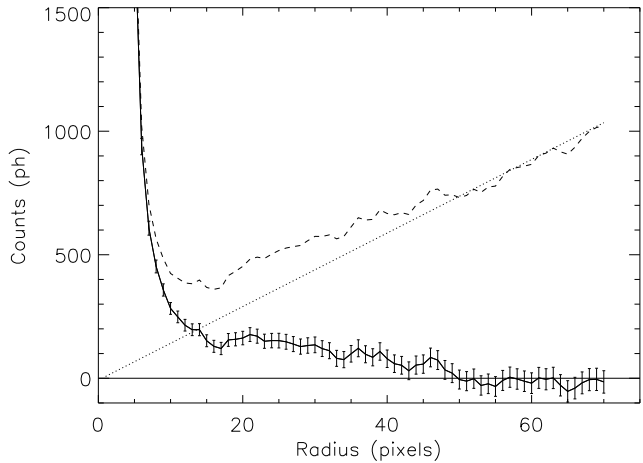}
\caption{Data for Hz2 in the b filter with a hardware window size of $960 \times 960$. The
  dashed line represents the number of counts in each consecutive ring for
  the source plus background measurements, the dotted line is the number of
  counts in each consecutive ring for the background only, and the solid black
  line is the number of counts in the rings once the background has been subtracted. The error bars on the background subtracted data are due to Poisson statistics.}
\label{FigEXT}
\end{figure}

For each case we measured the total number of source and background counts in 70
concentric rings around the source; each ring one pixel wide and varying in
radius from $1-70$ pixels from the centre of the source. An example is shown in
Figure~\ref{FigEXT}. The background subtracted source counts are represented
by the solid black line, and in this example the radius at which the total
number of counts becomes indistinguishable from the background, is at around 50
pixels radius. Table~\ref{tabRadius} shows the radius at which the count rate per pixel of
the source becomes indistinguishable from the background. These results show
that even in the case of the faint source, WD1121+145, very faint PSF wings are
still sometimes observed. 

\begin{table}
\caption{Radius at which the count rate per pixel of the source becomes
  indistinguishable from the background.}
\label{tabRadius}
\begin{tabular}{@{}lccc}
\hline
Filter & Hz2  & GD128 & WD1121+145  \\
& (pixels) & (pixels) & (pixels) \\
\hline
v &  20 & - & 15 \\
b & 50 & - & 15 \\ 
u & 51 & - & 21 \\
uvw1 & 52 & - & 45 \\
uvm2 & 54 & - & 52 \\
uvw2 & 55 & - & 48 \\
white & - & 30 & 20 \\
\hline
\end{tabular}
\end{table}

From these results the background region should be at least 55 pixels from the
centre of the source, thereby avoiding any PSF wing photons in the background 
calculations. We used an annulus with inner radius 55 pixels from the centre of
the source and an outer radius of 70 pixels to produce a large sample of
background area with which to calculate the background contribution.

\subsection{Count rate calculations}
\label{cr calc}
The method used to obtain corrected count rates for standard star UVOT
calibration observations is outlined in the following section.  This is the
procedure that was used for the calibration analysis, and is not necessarily a recipe for
analysing GRB data. Please go to Section~\ref{practical} for practical guidelines to analysing data.

The raw image data from the UVOT was preprocessed using the standard {\it
Swift} pipeline (UVOT2FITS v3.16).  The first step of the pipeline is to
construct a bad pixel map for each exposure and then to use this map to
exclude pixels from further analysis.  Next, fk5 equatorial coordinates
were applied to each exposure and the exposures were rotated so that
north is up and east to the left.  This coordinate system is applied
using information from the {\it Swift} star trackers, which are accurate to
five arcseconds. The pipeline corrects for this uncertainty by matching
the star field in each exposure to sources in the HST Guide Star
Catalogue.  This aspect correction is accurate to 0.5
arcseconds (90 percent confidence radius). The radial aperture size of 10 pixels meant that a fixed
pattern correction (see Section~\ref{tech}) did not need to be applied to these data.

Following is a list of steps taken to determine the count rates of standard stars
from the sky images produced by the {\it Swift} pipeline:

\begin{enumerate}
\item We obtained an aspect-corrected image in SKY coordinates from the {\it Swift} archive
  (http://swift.nasa.gov/docs/swift/archive).
\item We removed any exposures or observations that contained any exposure time
  anomalies\footnote{Please see 
  http://heasarc.gsfc.nasa.gov/docs/swift/analysis/uvot\_digest.html for more
  details on timing problems.}.
\item Where multiple exposures were taken in an observation, count rates were
  calculated for the individual exposures, and then a weighted mean obtained.
\item We obtained observed (source $+$ background) counts using a 10 pixel (5
  arcsec) radius aperture.
\item We obtained the dead-time corrected exposure time for each exposure to
  calculate the count rate (the EXPOSURE keyword in the UVOT sky image files).
\item We obtained the background count rate from an annulus with an inner radius 
of 55 pixels
  and an outer radius of 70 pixels centered on the source. If the
  background level was below 10 ph/pix we used a  mean background method, or if 
it was  above this level we used a clipped-mean background method, as described
in Section~\ref{back meth}.
\item We corrected the  (source $+$ background) and background  count rates for
  coincidence loss using Equation~\ref{Equ5}, as described in Section~\ref{coi}.
\item Finally, we subtracted the coincidence-loss-corrected background 
count rate from the coincidence-loss-corrected (source
  $+$ background) count rate to obtain the coincidence-loss-corrected 
source count rate.
\end{enumerate}

Errors on the count rates were calculated using Poisson statistics and were carried
through the coincidence loss equations to produce final coincidence-loss-corrected 
count rate
errors.

\section{Coincidence Loss}
\label{coi}
\begin{figure}
\includegraphics[angle=0,width=84mm]{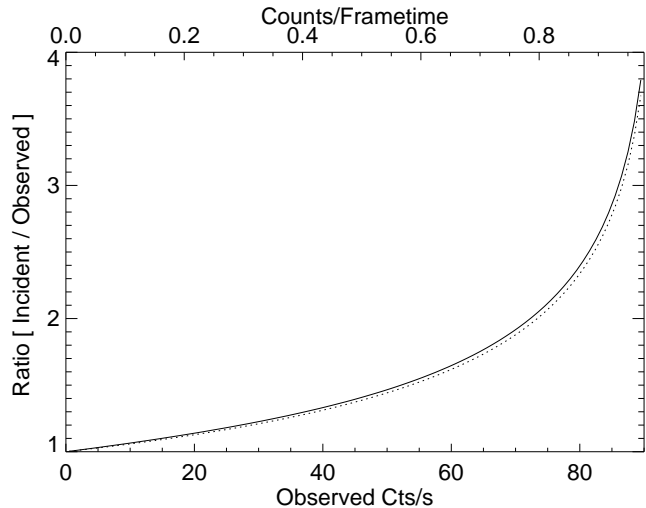}
\caption{The size of the coincidence loss correction (the ratio of the incident
  to observed counts) is shown as a function of the 
observed dead-time corrected count rate for the nominal full-frame observing 
mode.   The dotted line shows the expected relation for a single pixel device, 
while the solid line includes our empirical polynomial adjustment to the 
theoretical relation. The top axis shows the number of counts per frame.}
\label{FigWayne1}
\end{figure}

As described in Section~\ref{tech}, the UVOT detector suffers 
from coincidence losses at high count rate when
multiple photons arrive at the same location on the detector during a 
single frame (see Figure~\ref{FigWayne1}). The theoretical
coincidence-loss-corrected count rate for a single-pixel detector is:
\begin{equation}
C_{theory}=\frac{-\ln(1-\alpha C_{raw}f_t)}{\alpha f_t}
\label{Equ3}
\end{equation}
where $C_{theory}$ is the incident count rate (in counts per second) and $C_{raw}$ is the raw observed
count rate (also in counts per second) calculated using the {\it dead-time corrected} exposure 
time (keyword EXPOSURE in the UVOT image FITS file header). $f_t$ is the
frame time (0.0110329s for a full frame; keyword FRAMTIME in the FITS file header) 
and $\alpha$ is the dead time correction factor (one minus the dead time fraction;
0.9842 for a full frame; keyword DEADC in the FITS file header). 

This theoretical coincidence-loss expression cannot be
applied on a pixel-by-pixel basis to UVOT images because each count assigned
to a UVOT pixel is determined by centroiding a photon splash over five 
physical CCD pixels (in a `cross-hair' sampling). 
The effect of high coincidence loss is thus not only to lose
counts, but also to reposition counts, since overlapping photon splashes
within a single frame will be misplaced by the centroiding algorithm.

\begin{figure}
\includegraphics[angle=0,width=84mm]{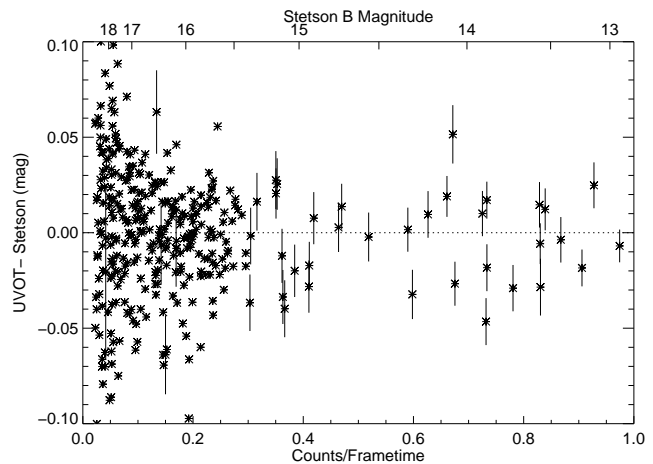}
\caption{The difference between UVOT and \citet{b20} magnitudes for 361 isolated
stars in the open cluster NGC 188 is plotted against the number of counts per
frame (bottom x-axis) and Stetson B magnitude (top x-axis). A polynomial
correction has been applied to the theoretical coincidence loss curve for UVOT 
to minimize any systematic trend with the Stetson magnitudes. For clarity, only 
selected Poisson error bars are included for counts per frametime of less than 0.3.}
\label{FigWayne2}
\end{figure}

We therefore introduce an empirical polynomial correction to account for the 
differences between the observed and theoretical coincidence loss 
correction:  
\begin{equation}
f(x)=1 + a_1 \times x + a_2 \times x^2 + a_3 \times x^3 + a_4 \times x^4
\label{Equ4}
\end{equation}
where $x=C_{raw}f_t$. Hence we have a full coincidence loss corrected incident count rate of:
\begin{equation}
C_{corr} = {C_{theory}}{f(x)}
\label{Equ5}
\end{equation}
where $C_{corr}$ and $C_{theory}$ are both in counts per second.
The coefficients in Equation~\ref{Equ4} were determined by a least-square fit to
minimize the differences between the  UVOT b magnitudes of 361 isolated stars 
in the open cluster NGC 188, with the catalogue of \citet{b20}. The count rates
were measured within 10 pixels as described in Section~\ref{cr calc}. The fit yielded 
the values $a_1 = 0.066$, $a_2 = -0.091$, $a_3 = 0.029$, and $a_4 = 0.031$.
Figure~\ref{FigWayne2} compares the UVOT and Stetson photometry of the 361
stars in NGC188 after the polynomial correction has been applied to the UVOT
data. This correction was verified using UVOT
observations of other  photometric fields (SA95, M67, SA104) also studied
by \citet{b21}. Note that the polynomial component changes the coincidence loss 
correction by less than 3 per cent for count rates up to 0.96 counts per frame 
(87 counts per second for a full frame), 
at which point the coincidence loss correction is a factor of 3.4.

The coincidence loss correction here applies to isolated point sources and does 
not apply for crowded or extended sources, or when a smaller aperture is needed 
to maximize the signal to noise ratio.
There is no general solution to these problems though we note that for faint 
sources, it may be preferable to maximize the signal to noise ratio with a
small aperture, and accept some increased scatter in the
small coincidence loss correction (see Section~\ref{practical}). 

The finite number of frames in an exposure implies that the measured count rate 
follows a Binomial distribution. Because of this, normal photometry packages
which assume Poisson statistics will not give an accurate measure of the error,
except at low count rates. Since the incident count rate must be derived by use 
of the non-linear coincidence-loss correction according 
to equation \ref{Equ3}, the error in the incident count rate,
$\sigma_{theory}$, is given by (see \citet{kuin07} ):

\begin{eqnarray}
\label{coierreq}
 {\sigma_{theory}^{\pm} = {-\frac{1}{\alpha f_t}}\  {ln(1 \pm {\frac{\sigma_{raw}
	f_t}{(1-C_{raw} f_t)}}) } }
\end{eqnarray}
 
where $\sigma_{theory}$ is in counts per second, and $\sigma_{raw} = \sqrt{C_{raw}
  (1-C_{raw} f_t)/t_{elapsed}}$ (in
  counts per second) is the binomial error in the measured count rate. For low 
count rates, the Poisson error is therefore a good approximation.
For the highest incoming photon fluxes (more than 0.9 counts per frame) the upper error becomes 
larger than the lower error, but in most cases of interest, they are nearly equal. 

Based on the NGC 188 photometry, any additional systematic error introduced by 
the coincidence loss correction is less than 0.01 magnitudes. This is confirmed 
in the comparison data sets where no trend is seen with magnitude.

\section{Position-Dependent Uniformity}
\label{uniform}
The photometric sources were all measured near the middle of the detector, but
any variation in sensitivity from position to position due to irregularities in
the fibre bundle, photocathode etc., or any larger scale trend in sensitivity
over the field of view clearly should be taken into account.
 
\begin{figure}
\includegraphics[angle=270,width=84mm]{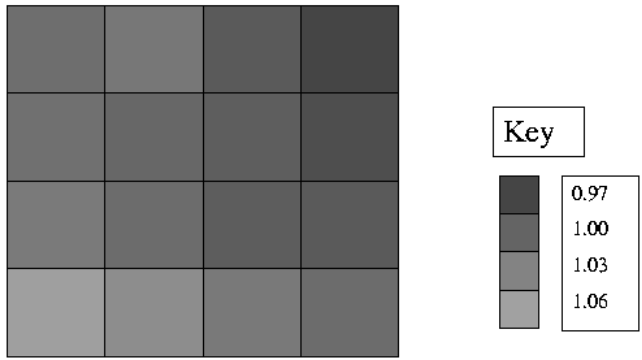}
\caption{Shaded plot showing how the sensitivity of the UVOT varies with
  position. The paler areas are more sensitive. The data are binned to make $4
  \times 4$ bins with a minimum of 4 points per bin. }
\label{FigAlice}
\end{figure}

Large scale sensitivity (LSS) variations over the area of the detector were
measured using the count rate of a standard star at a variety of positions on 
the detector. For this we used GD50, and several other stable stars to evenly
sample the whole field of view, giving 163 points in total. The standard
deviation of the normalised count rate of all these points is 3.2 per cent, but for
the central $1024\time 1024$ region only, the standard deviation is 2.0 per
cent. A trend of count rate versus position could clearly be seen, so the data were
binned spatially into 16 bins, with a minimum
of 4 measurements per bin, to remove small scale variations. The LSS variation over the
full field of view demonstrated by the binned data has a 2.2 per cent standard
deviation with a peak to trough range of 8 per cent.  
Figure~\ref{FigAlice} is a shaded plot showing this
variation. One corner has a higher sensitivity and the opposite corner
lower. However, after binning, the central region of the detector (the 
central $1024\time 1024$ pixels, equivalent to the middle 4 boxes in the
diagram) within which the photometry measurements were taken appears to
have more-or-less flat sensitivity showing a standard deviation of 0.6 per cent.

An upper limit to the small scale, or pixel to pixel sensitivity variation, is
found from repeated measurements of the same object (GD50) at slightly
different positions in the central area and is found to be at most 2.3 per cent
standard deviation. 
This has been confirmed by the measurements of several stars used in the long
term stability calibration, and also by the pixel-to-pixel variations in count
rate in exposures taken with the on board LED. At present, it would be prudent
to assume a 2.3 percent systematic error on individual UVOT photometry measurements.

A more detailed analysis of large and small scale
sensitivity and of the sensitivity changes over the lifetime of UVOT 
will be presented in a future paper~\citep{bALICE}.

\section{In-Orbit Effective Area Determination}
\label{eff area}
The predicted effective area curves for each filter were calculated in units of cm$^{2}$
by multiplying the UVOT reference instrument response (Figure~\ref{figInstResp}) 
by the ground based filter transmission curves (Figure~\ref{FigFT}). This section describes
how the curves were adjusted to produce the in-orbit effective area curves.

\subsection{In-orbit instrument response curve}
To obtain in-orbit effective area curves, the instrument response curve was
adjusted using an in-orbit correction curve. The correction curve was produced 
using the standard 
star observations through the UV and optical filters. For each star, in each filter,
the ratio of the observed count rate to the predicted count rate was
computed. These ratios 
were averaged for all the stars in each filter. These average 
in-orbit/predicted ratios and their corresponding
standard errors can be seen in Table~\ref{tabRatio}.

\begin{table}
\caption{Ratio of observed to predicted count rates used to produce an in-orbit correction curve.}
\label{tabRatio}
\begin{tabular}{@{}lccc}
\hline
Filter & Average count  & Standard & Number of \\
 & rate ratio & Error & Observations \\
\hline
v & 0.698 & 0.008 & 13 \\
b & 0.725 & 0.011 & 12 \\
u & 0.753 & 0.014 & 11 \\
uvw1 & 0.786 & 0.009 & 3 \\
uvm2 & 0.740 & 0.010 & 3 \\
uvw2 & 0.784 & 0.009 & 3 \\
\hline
\end{tabular}
\end{table}

The correction curve was produced by fixing the ratio values at the
central wavelengths of each filter (Table~\ref{tabPete}), and applying a spline
fit. The correction curve was assumed to be constant longwards of the central 
wavelength of the v filter and shortwards of the central 
wavelength of the uvw2 filter.
Figure~\ref{FigCC} shows the in-orbit correction
curve. The shape of this curve, which represents the deviation from the ideal response, is not consistent with significant contamination of the optical surfaces by molecular material, which would have a greater effect in the UV than in the
optical. This in-orbit correction curve was then multiplied by the reference instrument
response curve to produce the in-orbit instrument response curve shown as the
solid black line in Figure~{\ref{figInstResp}.

\begin{figure}
\includegraphics[angle=90,width=84mm]{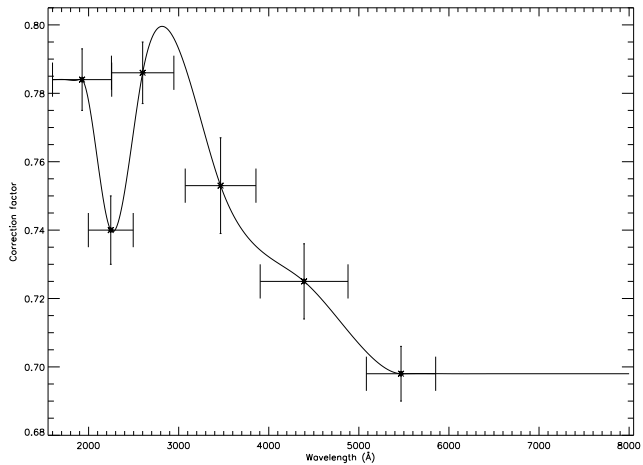}
\caption{In-orbit correction curve made by a spline fit to the ratios given in Table~\ref{tabRatio}. The y-axis error bars show the standard
  error for each ratio; the x-axis error bars show the FWHM of each filter
  (Table~\ref{tabPete}).} 
\label{FigCC}
\end{figure}

\subsection{In-orbit effective area curves}
\label{in-orbit eff}
The in-orbit effective area curve for each filter, except for white, 
was calculated by multiplying the in-orbit instrument response with 
the ground-based filter transmission curves. Figure~\ref{figEffArea} 
shows the resulting in-orbit effective area curves.

The white filter is a special case because it covers a very wide band, 
and so the shape of its effective area curve depends on the shape of the instrument 
response curve throughout the wavelength range. Unfortunately, it is also 
the one filter that does not have a well measured throughput, and so a 
different sequence was followed for the calibration of the white filter 
effective area.  First, we combined the white filter predicted transmission 
curve with the in-orbit instrument response function to predict the count 
rates of the standard stars observed through this filter. We then compared 
these with the observed count rates, to find the average in-orbit correction 
factor for the white filter. This ratio was found to be $0.88 \pm 0.05$. 
Unlike the other optical and UV filters, the white filter in-orbit effective 
area curve was then calculated by combining the in-orbit instrument response 
curve with the theoretical white filter transmission curve and then multiplying 
by the extra factor of 0.88.

\begin{figure}
\includegraphics[angle=90,width=84mm]{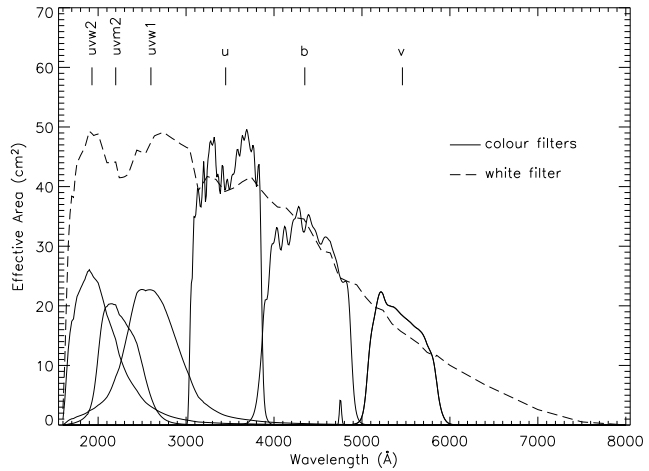}
\caption{In-orbit effective area curves for the UVOT filters.}
\label{figEffArea}
\end{figure}

\section{Photometric Zero Points}
\label{zp}

The photometric zero point $Z_{pt}$ of each UVOT filter is defined as the magnitude 
which is equivalent to 1
count per second \citep{b13}, and is given by the equation:
\begin{equation}
Z_{pt}=m_{source}+2.5\log\left(C_{source}\right),
\end{equation}
where $m_{source}$ is the magnitude of a source, and $C_{source}$ is the count
rate of an observed source after correction for coincidence loss and dead time.

We have taken the observed spectrum of Vega from \citet{b4} to define our UVOT magnitude
system. Thus this spectrum of Vega represents $m_{Vega}=0$ in all filters. 
Transformations are provided in Section~\ref{colour trans} to convert from the UVOT system to the Johnson system.  
The zero points for the optical and UV filters 
were calculated by standardising the count rates to
the Vega spectrum. The expected count rate of each
observed star (${\it C_{exp}}(i)$) was calculated by folding its 
spectrum through the in-orbit filter effective areas (Section~\ref{in-orbit eff}). In the same way the spectrum of Vega was
used to produce an expected Vega count rate
($C_{exp}(vega)$). The zero points ($Z_{pt}(i)$) for each source in
each filter were then calculated using:
\begin{equation}
Z_{pt}(i) = m_{Vega} + 2.5\log\left(C_{exp}(vega)\frac{C_{obs}(i)}{C_{exp}(i)}\right).
\label{Equ2}
\end{equation}
The final zero point ($Z_{pt}$) for each filter was
calculated by averaging over all the observations in that
filter.  

Figure~\ref{figZeropts}
shows the data used to produce the zero
points for each of the UVOT filters. The final, mean zero point is shown 
with a dashed line, and the RMS scatter is illustrated with dotted lines to
either side. The error bars shown on the individual points in the plots 
include the Poisson error in the
raw observed count rate, and the errors associated with the
stellar spectra used. The predicted errors on the Landolt stars in the optical
filters were calculated using an estimate of the systematic error between the
Landolt and Johnson system \citep{b7} added in quadrature with the Landolt
colour term errors \citep{b1}. The predicted errors on the Landolt stars in the
white filter were calculated to be 4 per cent due to the scatter of possible
Pickles spectral matches and this was added in quadrature with the Landolt
colour term errors \citep{b1}. The Oke and white dwarf predicted errors in all
filters were calculated using a HST spectrophotometric error of 2 per cent 
\citep{b11} or an IUE error of
3 per cent \citep{b12}. 

Table~\ref{tabZeropts} shows the final zero points. For each
filter the average error, which 
is the mean of the individual errors for each point, is given in
Column 6.  As expected, this average error is comparable to
the RMS scatter about the mean, which is given in Column 5. The error
shown in Column 4 is the standard error for each zero
point, which is a measure of the error on the mean and is smaller than the
RMS. However, the standard errors for the UV zeropoints are based on only three
data points in each filter, and the standard error for the white zeropoint is
based on only six data points, so these cannot be considered as valid estimates
of the uncertainties on the zeropoints.
 For
this reason, for the UV and white zero points, we recommend using the errors
from column 6 in Table~\ref{tabZeropts} as the uncertainties on the
zeropoints. For the convenience of the reader we have listed the 
recommended zeropoint error for all filters in Column 3.

\begin{table}
\caption{ In-orbit zero points. The standard error (Column 4) is the error on the mean 
zero point for each filter; the RMS (Column 5) gives the scatter; the average
error (Column 6) 
includes uncertainties on individual measurements. Column 3 (Recommended Uncertainty) 
lists the recommended error for each zero point.}
\label{tabZeropts}
\begin{tabular}{@{}lccccc}
\hline
Filter & Zero Point & Recommended & Standard & RMS & Average \\
 & & Uncertainty & Error & & Error \\ 
\hline
v & 17.89 & 0.013 & 0.013 & 0.04 & 0.03 \\
b & 19.11 & 0.016 & 0.016 & 0.05 & 0.05 \\
u & 18.34 & 0.020 & 0.020 & 0.06 & 0.07 \\
uvw1 & 17.49 & 0.03 & 0.013 & 0.02 & 0.03\\
uvm2 & 16.82 & 0.03 & 0.015 & 0.02 & 0.03\\
uvw2 & 17.35 & 0.03 & 0.012 & 0.02 & 0.03\\
white & 20.29 & 0.04 & 0.023 & 0.05 & 0.04\\
\hline
\end{tabular}
\end{table}

\begin{figure*}
\includegraphics[width=160mm]{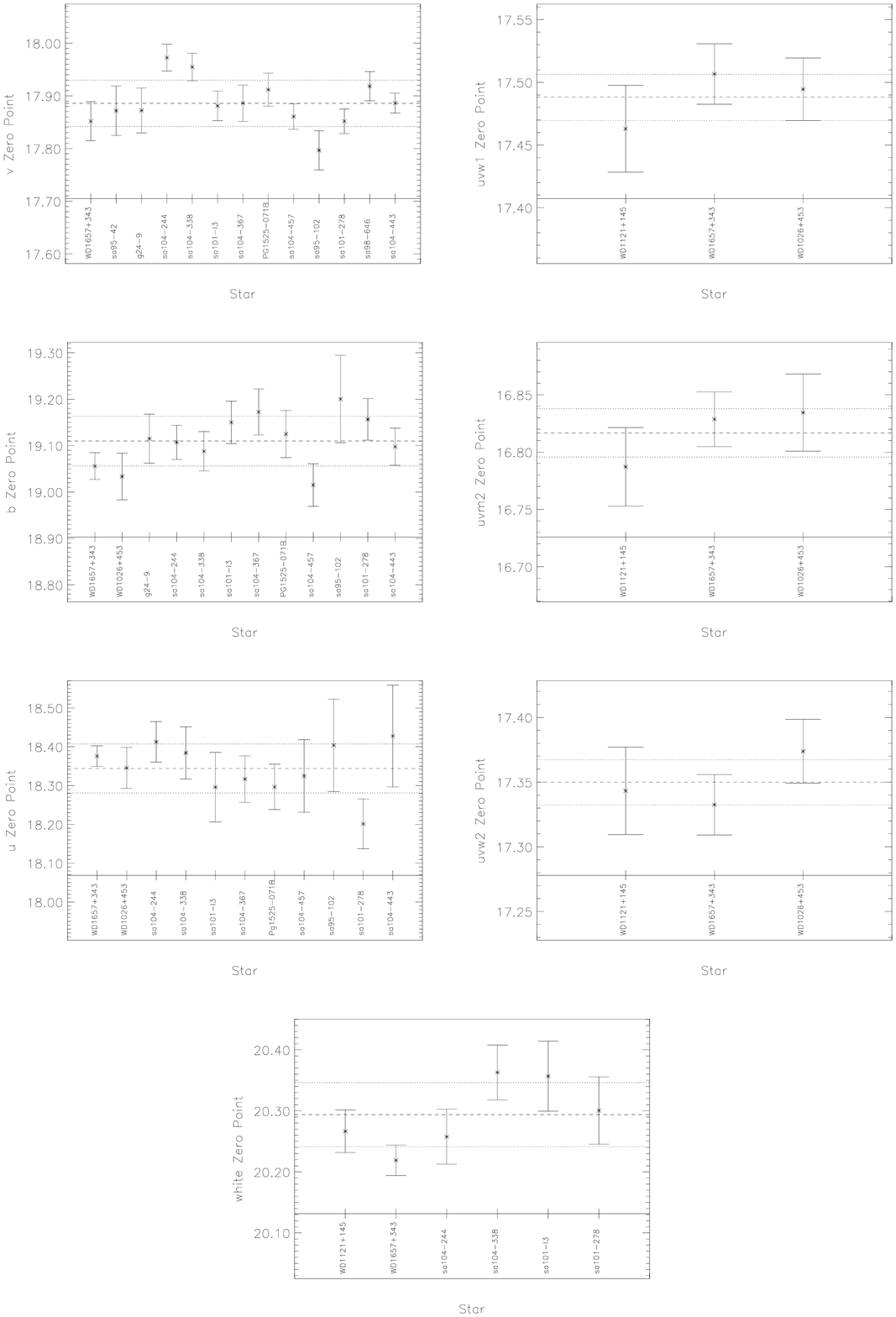}
\caption{Zero point calculations for the UVOT filters. Left column: v, b and u
filters; right column: uvw1, uvm2 and uvw2 filters; bottom: white filter. Each point represents the zero point obtained from a single standard star, labelled individually on the x-axis. The error bars include the Poisson error in the raw observed count rate and the
errors associated with the stellar spectra used (see Section~\ref{zp}). 
The dashed line indicates the mean zero point, with the dotted lines
showing the 1 sigma RMS error.}
\label{figZeropts}
\end{figure*}

\section{Colour Transformations}
\label{colour trans}
Colour transformations are needed to convert from the UVOT magnitude
system to any other system, in order to compare
UVOT data with photometry from other sources. In this section we provide colour transformations to convert
from the UVOT system to the Johnson system for a range of stellar spectra.We
also provide transformations for a set of synthetic GRB spectra because {\it
  Swift} is primarily a GRB mission.

The colour transforms from the UVOT ubv system to the Johnson UBV system
were calculated for stars using Pickles spectra~\citep{b2}, and for
GRB model spectra. The model GRB spectra were generated assuming a power law 
continuum and varying degrees of dust extinction, with the form 
$F_{\nu}^{-\beta} exp^{-\varepsilon}$,
where $\beta$ is the GRB spectral index, and ${\varepsilon}=A_{\lambda}/{A_v}$ 
is the relative exinction per unit wavelength. The wavelength dependence of the 
extinction was modelled on the SMC extinction law (Pei 1992), and GRB spectra 
were produced for $\beta=0,1,2$, rest-frame visual extinctions 
$A_v=0.0,0.2,0.4,0.8,1.0$~mag, and redshift in the range $0.3 < z < 2.0$.

To produce the Johnson colours and magnitudes, the Pickles spectra
and GRB models were folded through the Johnson response
curves (described in Section~\ref{standards}). To produce the UVOT colours 
and magnitudes the same spectral models were folded through the UVOT in-orbit 
effective areas to produce expected UVOT count rates ($C_{source}$),
then converted into magnitudes using:
\begin{equation}
m_{source}=Z_{pt}-2.5\log(C_{source})
\label{equm}
\end{equation}
where $m_{source}$ is the magnitude of the source in the UVOT system, and
$Z_{pt}$ is the zero point of each filter (listed in Table~\ref{tabZeropts}).

Figure~\ref{FigColor} plots the difference between UVOT and Johnson colours 
against UVOT colours for the optical filter combinations. The stars in 
each plot represent the Pickles stars, and the
triangles represent the GRB models. The
solid line in each plot shows the second order polynomial fit for Pickles stars;
the dotted line shows the second order polynomial fit for the GRB models, and the
dashed line shows where the colours in the two systems would be equal. The RMS error on the residuals, and the
ranges for which the fits have been calculated, can be seen
in Table~\ref{tabColRMS}. The colour terms in these Pickles star
polynomial fits are:
\break
\break
U-B=0.034[$\pm$0.007]+0.862[$\pm$0.007](u-b)\break
\vbox{\hspace{0.5cm}+0.055[$\pm$0.006](u-b)$^2$}
\break
B-V=-0.004[$\pm$0.004]+1.039[$\pm$0.011](b-v)\break
\vbox{\hspace{0.5cm}-0.037[$\pm$0.007](b-v)$^2$}
\break
U-V=0.071[$\pm$0.010]+0.899[$\pm$0.008](u-v)\break
\vbox{\hspace{0.5cm}+0.018[$\pm$0.003](u-v)$^2$}
\break

The colour terms calculated from the second order polynomial fits for the GRB
models are:
\break
\break
U-B=0.086[$\pm$0.003]+0.886[$\pm$0.007](u-b)\break
\vbox{\hspace{0.5cm}+0.050[$\pm$0.006](u-b)$^2$}
\break
B-V=-0.008[$\pm$0.001]+1.012[$\pm$0.003](b-v)\break
\vbox{\hspace{0.5cm}-0.018[$\pm$0.002](b-v)$^2$}
\break
U-V=0.162[$\pm$0.002]+0.904[$\pm$0.002](u-v)\break
\vbox{\hspace{0.5cm}+0.010[$\pm$0.002](u-v)$^2$}
\break
Again, the ranges of colours over which the transforms were calculated are
given in Columns 5 and 6 of Table~\ref{tabColRMS}.

\begin{figure}
\includegraphics[width=84mm]{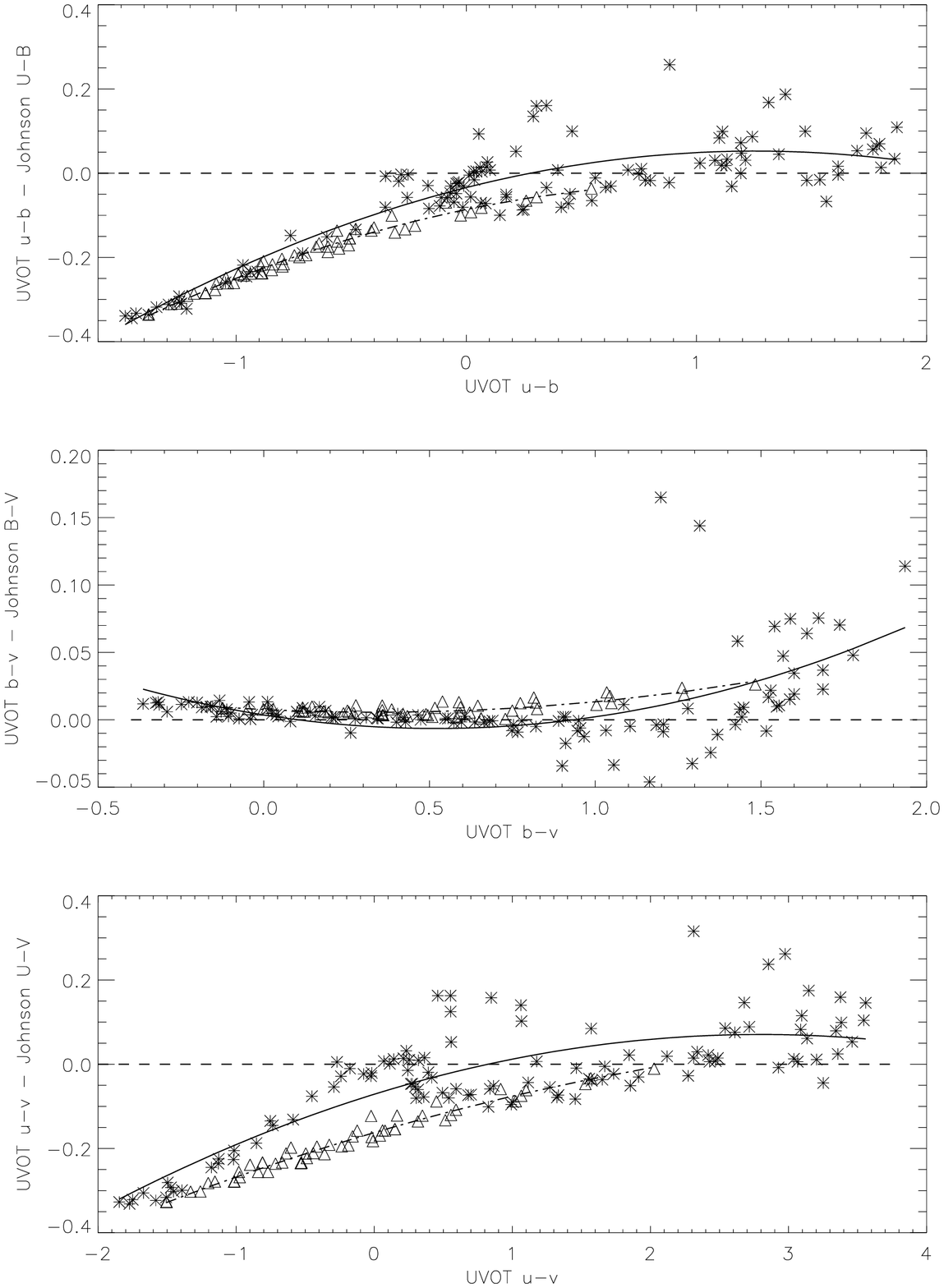}
\caption{The differences between UVOT optical colours and Johnson optical colours for Pickles stars (stars)
  and GRB models (triangles): UVOT u-b compared with Johnson U-B versus UVOT
  u-b (top  plot); UVOT b-v compared with Johnson B-V versus
  UVOT b-v (middle plot); and UVOT u-v compared with Johnson U-V versus UVOT u-v  (bottom plot).  
The solid lines represent the second order polynomial
  fits to the Pickles stars, the dotted lines represent the second order 
polynomial fit to the GRB models, and the dashed lines show where the 
colours in the two systems are equal. The polynomial fit parameters are given in
  Section~\ref{colour trans}.}
\label{FigColor}
\end{figure}

\begin{table}
\caption{RMS error on residuals to colour fits. PS is Pickles star data fits and
GRB is GRB models data fits. Columns 5 and 6 give the minimum and maximum limits
on the x-axis (of Figures~\ref{FigColor} and~\ref{FigJohn}) respectively,
which give the range of colours over which the transforms were calculated.}
\label{tabColRMS}
\begin{tabular}{@{}lccccc}
\hline
Fit  & y-axis & x-axis & RMS & Minimum & Maximum \\
 & & & error & x-axis & x-axis \\
\hline
PS & U-B & u-b & 0.057 & -1.482 & 1.871 \\
PS & B-V & b-v & 0.025 & -0.364 & 1.935 \\
PS & U-V & u-v & 0.075 & -1.846 & 3.558 \\
GRB & U-B & u-b & 0.078 & -1.380 & 0.543 \\
GRB & B-V & b-v & 0.034 & -0.124 & 1.483 \\
GRB & U-V & u-v & 0.102 & -1.505 & 2.026 \\
PS & B-b & b-v & 0.020 & -0.364 & 1.935 \\
PS & B-b & u-b & 0.030 & -1.482 & 1.871 \\
PS & V-v & b-v & 0.014 & -0.364 & 1.935 \\
PS & V-v & u-v & 0.015 & -1.846 & 3.558 \\
PS & U-u & u-b & 0.073 & -1.482 & 1.871 \\
PS & U-u & u-v & 0.071 & -1.846 & 3.558 \\
GRB & B-b & b-v & 0.001 & -0.124 & 1.483 \\
GRB & B-b & u-b & 0.002 & -1.380 & 0.543 \\
GRB & V-v & b-v & 0.004 & -0.124 & 1.483 \\
GRB & V-v & u-v & 0.004 & -1.505 & 2.026 \\
GRB & U-u & u-b & 0.011 & -1.380 & 0.543 \\
GRB & U-u & u-v & 0.012 & -1.505 & 2.026 \\
\hline
\end{tabular}
\end{table}

The observations of one faint white dwarf star and nine
Landolt stars were then compared with these transforms. Observed count rates
($C_{obs}$) 
were obtained using the method
described in Section~\ref{cr calc}, and converted into magnitudes using Equation~\ref{equm}.
Figure~\ref{Figcolobs} plots the  Pickles star fits to the Johnson versus UVOT
colours together with these
observational data. The error bars on the observed data show the Poisson error
in the raw observed count rate on the x-axis, and the errors associated with the
Landolt colour terms in the y-axis \citep{b1}. This Figure shows that within 
the scatter of these observations,
the fits produced with the Pickles stars agree with the observations.

\begin{figure}
\includegraphics[width=84mm]{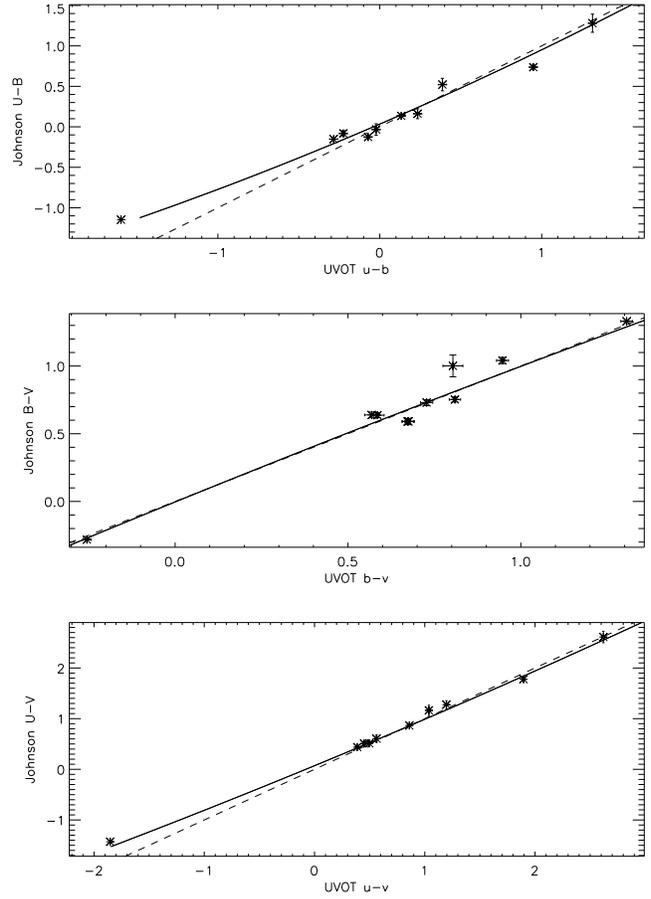}
\caption{Johnson optical colours versus UVOT optical colours using observed
  data. The solid lines are the polynomial fit to the Pickles spectra
  shown in Figure~\ref{FigColor} and described in Section~\ref{colour trans}. 
The dashed  lines show where the colours in the Johnson and UVOT systems are equal.}
\label{Figcolobs}
\end{figure}

Figure~\ref{FigJohn} plots the difference between Johnson and 
UVOT magnitudes against UVOT colours.  
The stars in each plot represent the Pickles
stars, whereas the triangles represent the GRB models.  Figure~\ref{FigJohn}
shows that the Pickles spectra and GRB models follow different curves: the solid line
in each plot shows the third order polynomial fit to
the Pickles stars, and the dashed line shows the second order polynomial fit to
the GRB models. The residuals to the fits are shown in
the lower panel of each plot, and show good agreement within 0.05 magnitudes,
apart from a few outliers. As in Figure~\ref{FigColor}, the outliers are due to
Pickles spectra from stars with deep absorption features; we have included
these in the plots to give the 
reader an idea of the appropriateness of the colour transformations 
for a range of star types.  The RMS error on the
residuals can be seen in Table~\ref{tabColRMS}. The upper and lower colour 
limits for each colour transform are given
in Columns 5 and 6. 
The colour terms obtained from
the Pickles polynomial fits are:
\break
\break
B-b=0.021[$\pm$0.003]+0.005[$\pm$0.012](b-v)\break
\vbox{\hspace{0.5cm}-0.014[$\pm$0.022](b-v)$^2$-0.011[$\pm$0.010](b-v)$^3$}
\break
B-b=0.011[$\pm$0.004]-0.011[$\pm$0.008](u-b)\break
\vbox{\hspace{0.5cm}-0.008[$\pm$0.004](u-b)$^2$-0.002[$\pm$0.004](u-b)$^3$}
\break
V-v=0.029[$\pm$0.002]-0.009[$\pm$0.009](b-v)\break
\vbox{\hspace{0.5cm}-0.037[$\pm$0.016](b-v)$^2$+0.017[$\pm$0.007](b-v)$^3$}
\break
V-v=0.026[$\pm$0.002]-0.014[$\pm$0.002](u-v)\break
\vbox{\hspace{0.5cm}-0.005[$\pm$0.001](u-v)$^2$+0.002[$\pm$0.0005](u-v)$^3$}
\break
U-u=0.042[$\pm$0.010]-0.130[$\pm$0.020](u-b)\break
\vbox{\hspace{0.5cm}+0.053[$\pm$0.010](u-b)$^2$-0.013[$\pm$0.010](u-b)$^3$}
\break
U-u=0.069[$\pm$0.012]-0.093[$\pm$0.009](u-v)\break
\vbox{\hspace{0.5cm}+0.037[$\pm$0.007](u-v)$^2$-0.007[$\pm$0.002](u-v)$^3$}
\break
The colour terms obtained for the GRB model fits are:
\break
\break
B-b=0.016[$\pm$0.0003]-0.009[$\pm$0.001](b-v)\break
\vbox{\hspace{0.5cm}-0.023[$\pm$0.001](b-v)$^2$}
\break
B-b=-0.018[$\pm$0.0005]-0.045[$\pm$0.001](u-b)\break
\vbox{\hspace{0.5cm}-0.014[$\pm$0.001](u-b)$^2$}
\break
V-v=0.023[$\pm$0.001]-0.021[$\pm$0.003](b-v)\break
\vbox{\hspace{0.5cm}-0.005[$\pm$0.003](b-v)$^2$}
\break
V-v=0.010[$\pm$0.0007]-0.012[$\pm$0.0006](u-v)\break
\vbox{\hspace{0.5cm}-0.0009[$\pm$0.0006](u-v)$^2$}
\break
U-u=0.068[$\pm$0.003]-0.159[$\pm$0.007](u-b)\break
\vbox{\hspace{0.5cm}+0.036[$\pm$0.006](u-b)$^2$}
\break
U-u=0.172[$\pm$0.002]-0.108[$\pm$0.002](u-v)\break
\vbox{\hspace{0.5cm}+0.009[$\pm$0.002](u-v)$^2$}

\begin{figure*}
\includegraphics[width=160mm]{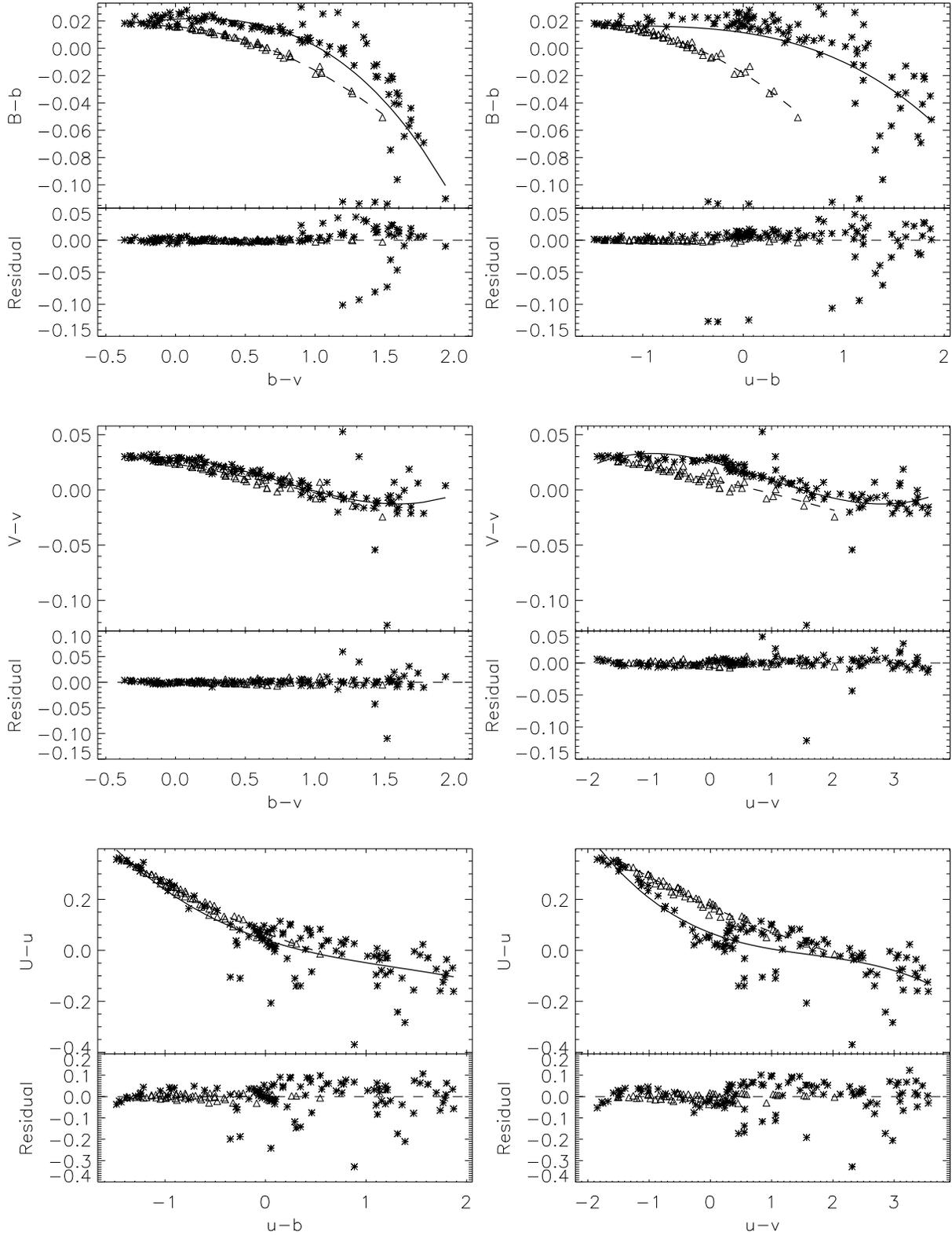}
\caption{The difference between Johnson and UVOT magnitudes versus UVOT optical
  colours
  for Pickles spectra (stars) and GRB models (triangles): B-b versus b-v 
(top left) and u-b (top
  right); V-v versus b-v (middle left) and u-v (middle right); U-u
  versus u-b (bottom left) and u-v (bottom right). The solid
  lines represent the third order polynomial  fits to the Pickles stars, and
  the dashed lines represent the second order polynomial fits to the GRB
  models. The polynomial parameters are given in Section~\ref{colour trans}. }
\label{FigJohn}
\end{figure*}

Figure~\ref{Figjohnobs} plots the difference between the Johnson and UVOT
magnitude versus UVOT colour for Pickles
star fits in comparison with observed data. As before, the error bars on the observed
data include the Poisson error
in the raw observed count rate, and the errors associated with the
Landolt colour terms \citep{b1}. 

\begin{figure*}
\includegraphics[width=160mm]{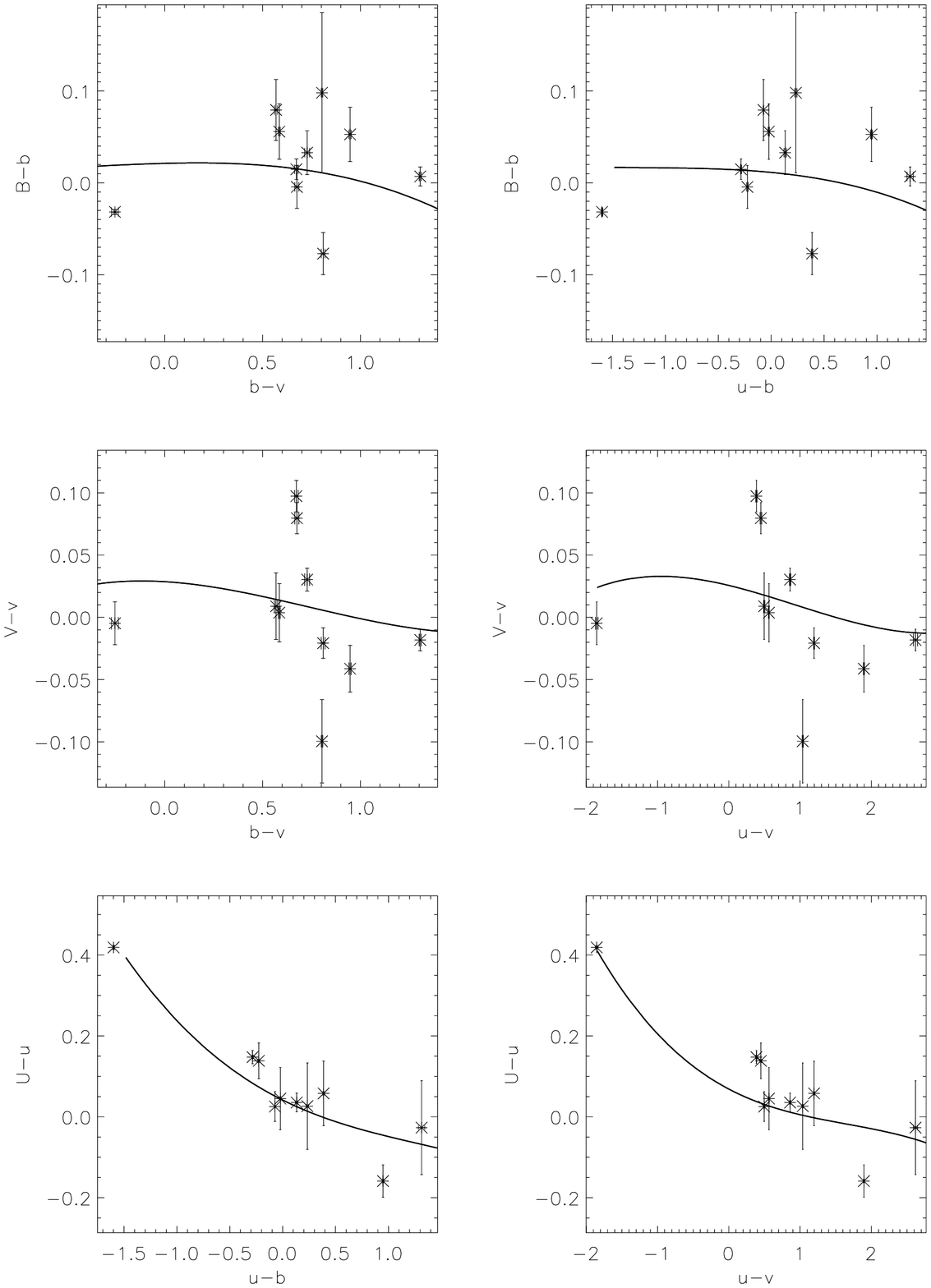}
\caption{The difference between Johnson and UVOT magnitudes versus UVOT optical
  colours for observed data. The solid line shows for comparison the polynomial
  fit to the Pickles spectra illustrated in Figure~\ref{FigJohn} and described
  in Section~\ref{colour trans}.}
\label{Figjohnobs}
\end{figure*}

\section{Count Rate to Flux Conversion}
\label{flux conv}
To compute an accurate flux density it is necessary to fold the source spectrum
through the effective area curves. However, we have found for all but the white
filter that there is not a strong dependency on stellar spectrum across a wide
range of classes. Thus, in many cases an estimate of the flux can be 
obtained directly from the count rate and this can be particularly useful when 
plotting UVOT data with data from other instruments. Therefore a count rate to flux
conversion factor has been calculated for each filter. Particular care should
be taken where there may be significant absorption or emission features in the
wavelength range of the filter (Table~\ref{tabPete}), and also with the
conversion factor for the white filter, which depends strongly on spectral
shape because of the wide wavelength range (1600--8000\AA).

The count rate to flux conversion for each filter was calculated, as described next, using Pickles
spectra~\citep{b2}, and a subset of the GRB power law spectral models
described in Section~\ref{colour trans} with redshifts ranging from $0.3 < z <
1.0$.

The effective wavelength ($\lambda_{eff}$) for each
filter was calculated using the Vega spectrum~\citep{b4} in the following weighted formula:
\begin{equation}
\lambda_{eff} = \frac{\int F_{vega}(\lambda)E_{area}(\lambda)\lambda d\lambda}{\int F_{vega}(\lambda)E_{area}(\lambda)d\lambda} 
\label{effwave}
\end{equation}
where $ F_{vega}(\lambda)$ is the Vega flux at a given wavelength, $\lambda$, 
$E_{area}(\lambda)$ is the predicted effective area.  The resultant
effective wavelengths are shown in Table~\ref{tabEffwave}. It must be noted 
that tails of the UV filter transmission curves extend into the optical range, 
as shown in Figure~\ref{FigFT}, and thus the effective wavelengths for the UV 
filters for very red spectra will be longer than those given in
Table~\ref{tabEffwave}.

A UVOT expected in-orbit count rate was calculated for each model spectrum in
each filter by folding the spectrum through the UVOT in-orbit effective
area curves (Section~\ref{in-orbit eff}). A flux value at the effective
wavelength for each model spectrum in each filter was
obtained by smoothing the spectrum into 10\AA~bins to remove narrow spectral
features, and then interpolating over four points around the effective
wavelength.

\begin{table}
\caption{Effective wavelengths for each filter, for a vega-like spectrum, as
  calculated using Equation~\ref{effwave}.}
\label{tabEffwave}
\begin{tabular}{@{}lc}
\hline
Filter & Wavelength (\AA) \\
\hline
v    & 5402 \\
b    & 4329 \\
u    & 3501 \\
uvw1 & 2634 \\
uvm2 & 2231 \\
uvw2 & 2030 \\ 
white & 3471\\
\hline
\end{tabular}
\end{table}

A count rate to flux conversion factor was then calculated for each spectrum
and averaged to produce a count rate to flux factor for each
filter.  The average count rate to flux conversion factor for the Pickles star spectra and
GRB models can be seen in Tables~\ref{tabCrPick} and~\ref{tabCrGRB}
respectively.  The RMS error (a measure of the data scatter) on the average
factor is also given in
each table, along with the range of UVOT b-v colours over which the factors
were calculated. Outside these ranges the conversion factors may not be applicable.

\begin{table}
\caption{Average count rate to flux conversion factor results for the Pickles
  star spectra. The units on the conversion factor and errors are ergs~cm$^{-2}$ s$^{-1}$ \AA$^{-1}$.}
\label{tabCrPick}
\begin{tabular}{@{}lcccc}
\hline
Filter & Conversion & RMS & Minimum & Maximum \\ 
 & factor & & b-v (mag) & b-v (mag) \\
\hline
v    &  $2.61\times10^{-16}$ & $2.4\times10^{-18}$ & -0.36 & 1.09\\
b    &  $1.32\times10^{-16}$ & $9.2\times10^{-18}$ & -0.36 & 1.09\\
u    &  $1.5\times10^{-16}$ & $1.4\times10^{-17}$ & -0.36 & 1.09\\
uvw1 &  $4.3\times10^{-16}$ & $2.1\times10^{-17}$ & -0.36 & 0.1\\
uvm2 &  $7.5\times10^{-16}$ & $1.1\times10^{-16}$ & -0.36 & 0.1\\
uvw2 &  $6.0\times10^{-16}$ & $6.4\times10^{-17}$ & -0.36 & 0.1\\
white & $2.7\times10^{-17}$ & $7.9\times10^{-18}$ & -0.36 & 1.09\\
\hline
\end{tabular}
\end{table}

\begin{table}
\caption{Average count rate to flux conversion factor results for GRB models. 
The units on the conversion factors and errors are ergs~cm$^{-2}$ s$^{-1}$ \AA$^{-1}$.}
\label{tabCrGRB}
\begin{tabular}{@{}lcccc}
\hline
Filter & Conversion & RMS & Minimum & Maximum \\
 &  factor & &  b-v & b-v \\ 
\hline
v    & $2.614\times10^{-16}$ & $8.7\times10^{-19}$ & -0.12 & 0.73\\
b    & $1.472\times10^{-16}$ & $5.7\times10^{-19}$ & -0.12 & 0.73\\
u    & $1.63\times10^{-16}$ & $2.5\times10^{-18}$ & -0.12 & 0.73\\
uvw1 & $4.00\times10^{-16}$ & $9.7\times10^{-18}$ & -0.12 & 0.03\\
uvm2 & $8.50\times10^{-16}$ & $5.6\times10^{-18}$ & -0.12 & 0.03\\
uvw2 & $6.2\times10^{-16}$ & $1.4\times10^{-17}$ & -0.12 & 0.03\\
white & $3.7\times10^{-17}$ & $4.9\times10^{-18}$ & -0.12 & 0.73\\
\hline
\end{tabular}
\end{table}

The large error in the white filter factor is due to large differences in the
convolution of blue and red spectra with the white filter wavelength range. 
The white filter factor for Pickles spectra and GRB models across the UVOT colour
$b-v$ and $uvw2-v$ are shown in Figures~\ref{FigWhite1} and~\ref{FigWhite2} respectively,
demonstrating the large scatter in the factor. In both figures the stars
represent the Pickles stars and the triangles represent the GRB models.

\begin{figure}
\includegraphics[angle=90,width=84mm]{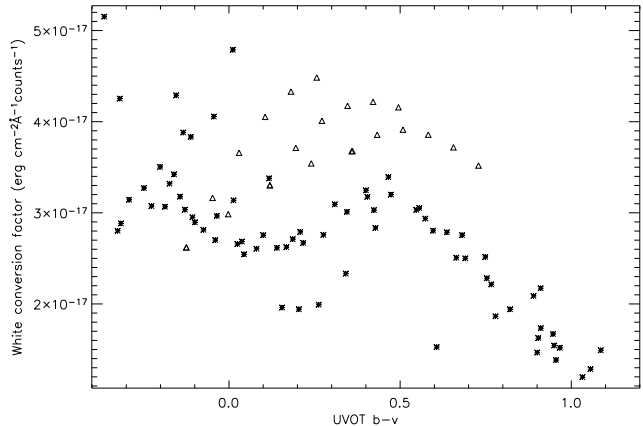} 
\caption{This plot shows how the white filter count rate to flux conversion factor varies with b-v UVOT colours.  The stars
represent the Pickles data, and the triangles represent the GRB model data.}
\label{FigWhite1}
\end{figure}

\begin{figure}
\includegraphics[angle=90,width=84mm]{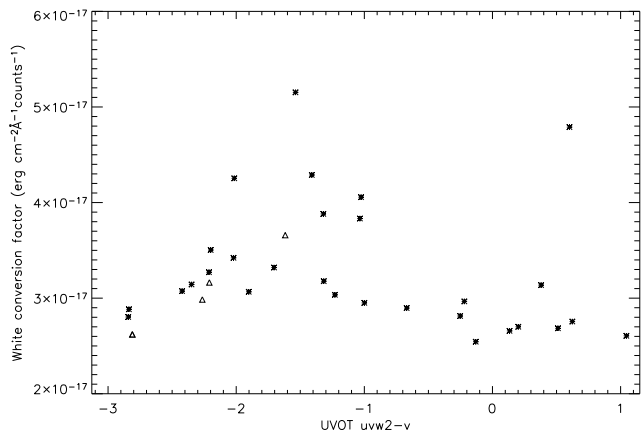}
\caption{This plot shows how the white filter count rate to flux conversion factor varies with uvw2-v UVOT colours.  The stars
represent the Pickles data, and the triangles represent the GRB model 
data.}
\label{FigWhite2}
\end{figure}

\section{Verification}
\label{verification}

We compared UVOT optical photometry measurements with photometry from other instruments 
in order to assess the accuracy of 
the UVOT photometric system independently. Three different groups working
independently, have performed the
comparison using different data, and using software other than that used for the
calibration work. The software used included the
latest {\it HEAsoft Swift} {\tt uvotsource} program in one case, and a combination of the 
{\tt uvotdetect} and {\tt uvotmag} (whose functionality has since been
taken over by {\tt uvotflux}) programs in another, while a test version of the CalDB 
incorporating the new calibration was employed. 

The tests directly compared measured optical magnitudes
with those published in the literature.
The sources used for u, b and v were: two sets of observations of 
23 stars in the GRB051021a field and 36 stars in the GRB051022 field (from
Henden ftp://ftp.aasvo.org/public/grb); 10 stars in the field of the
SN2005am (as calibrated by \citet{LW2006}); six sources in the GRB060218
field (calibrated by \citet{hick}). In addition the photometry of stars in the
field PG1633+099B have been compared with magnitudes given by
\citet{b21}; this work is shown in Section~\ref{example}.

Not all the measurements were made in the central region where the throughput
is most uniform (see Section~\ref{uniform}). Those in the GRB060218 and
SN2005am fields were distributed fairly randomly across the detector in
individual eposures. For the 
GRB051021a and GRB051022 fields, about half the
stars fell outside the central 4 squares in Figure~\ref{FigAlice}, but did not
fall in the corners; the large scale sensitivity for the positions of the 
majority of these sources is estimated to vary from 0.99 to 1.02.

The u, b, and v magnitudes were obtained from UVOT observations using
the HEAsoft tools mentioned above. After converting these instrumental magnitudes 
into Johnson U, B and V magnitudes using the colour
corrections (equations in Section~\ref{colour trans}), comparisons were made
with published U, B, V photometry. 

In general, the differences in magnitudes as observed by UVOT and those 
obtained from references as described above are small (see
Table~\ref{paulstable}). 
There is no evidence for a bias in any of the magnitude
differences.  The transformed UVOT colours $(b-v)_{trans}$
and $(u-b)_{trans}$ also match well with the reference samples. 

\begin{table}
\caption{Comparisons of UVOT measurements of optical magnitudes with literature
  magnitudes. The UVOT optical magnitudes and colours have been transformed to the
  Johnson system using the colour transformations.}

\label{paulstable}
\begin{tabular}{@{}lccccc}
\hline
magnitude & mean        & rms  & no.     & no.  \\
UVOT-other& difference  & error& sources & data \\
          & (mag)       &      &         & sets \\
\hline
$v_{trans} - V$         & 0.019 & 0.020 &81& 5 \\
$b_{trans} - B$         & -0.024 & 0.016 &81& 5 \\
$u_{trans} - U$         & -0.036 & 0.036 &65& 4 \\
$(b-v)_{trans} - (B-V)$ & -0.028 & 0.004  &59& 2 \\
$(u-b)_{trans} - (U-B)$ & 0.024  & 0.007  &59& 2 \\
%\hline
%$uvw1 - UVW1$   & 0.052  & 0.019  & 7& 2 \\
%$uvm2 - UVM2$   & -0.020 & 0.029 & 7& 2 \\
%$uvw2 - UVW2$   & 0.008  & 0.030 & 7& 2 \\
\hline
\end{tabular}
\end{table}

\subsection{Binned data}
\label{binning}
The photometric calibration of the UVOT is based upon
data taken in $1 \times 1$ binned (i.e. unbinned) modes.  However, a large
fraction of observations by the UVOT are made using $2 \times 2$ binned modes,
to reduce the telemetry volume generated on board the spacecraft.  We have
checked whether the binning mode has a significant effect on the photometric
calibration, by comparing the count rates of a set of stars observed in $ 1
\times 1$ and $2 \times 2$ binned modes.  The average difference in the count
rates for 30 stars in the field of GRB060206, in the UVOT v band using a 5
arcsec radius aperture (5 pixels for the $2 \times 2$ binning), 
is less than 1 per cent of the average count rate, which
is less than the average measurement error based on Poisson statistics.  We
conclude that there is no evidence for a significant difference in the count
rates measured in $1 \times 1$ and $2 \times 2$ binned modes. Thus this
calibration should be relevant to $2 \times 2$ binned and unbinned data.

\section{Practical photometry}
\label{practical}

The data analysis described earlier, in Section~\ref{photometry}, 
concerned the method used to obtain an optimal calibration over a 
broad range of source strengths. If the reader were to follow this method 
(using a 5 arcsec aperture and a large source-free background annulus more than 27 
arcsec from the source), they should obtain reasonable results. 

However, when doing photometry on point sources, particularly faint ones, 
the size and shape of the extraction aperture needs to be optimized to take into account the size
and shape of the PSF, the degree of crowding in the exposure, and the
desired science. For isolated point sources the optimal aperture should maximize the signal-to-noise
ratio. We have found that in most cases the maximum signal to noise is obtained
with an aperture radius between 2.5 and 4 arcsecs, but the actual value 
depends on the strength of the source, the density of sources in the field and the background level.
\citet{LW2006} similarly found an optimum aperture radius of 2.5 arcsec for unbinned UVOT  
data and 3 arcsec for binned data for the fields they were working with.  
Their work focussed on comparing UVOT optical photometry with standard fields, and
their calibration is based directly on these apertures.

\begin{table}
\caption{ Aperture corrections are given here in terms of magnitude for apertures 
having radii from 2 to 4.5 arcsec, calculated with the {\tt uvotapercorr} task
Swift\_Rel2.7(Bld21)\_15Jun2007, using the CalDB file
swureef20041120v102.fits. The 2.0 or 2.5 arcsec aperture radii are not generally
recommended, for reasons given in Section~\ref{practical}, but may be the best option 
where the field is very crowded or the background is complex.
}
\label{apercortable}
\begin{tabular}{@{}lcccccc}
\hline
Aperture \\
(arcsec): & 2.0 & 2.5 & 3.0 & 3.5 & 4.0 & 4.5 \\
\hline
v &    -0.276 & -0.145 & -0.091 &  -0.054 &  -0.032 &  -0.014 \\
b &    -0.327 & -0.176 & -0.111 &  -0.065 &  -0.037 &  -0.015\\
u &    -0.329 & -0.169 & -0.103 &  -0.059 &  -0.034 &  -0.015\\
uvw1 & -0.405 & -0.212 & -0.126 &  -0.069 &  -0.037 &  -0.015\\
uvm2 & -0.342 & -0.182 & -0.109 &  -0.060 &  -0.033 &  -0.014\\
uvw2 & -0.417 & -0.222 & -0.133 &  -0.073 &  -0.039 &  -0.016\\
white (b) & -0.327 & -0.176 & -0.111 &  -0.065 &  -0.037 &  -0.015\\
\hline
\end{tabular}
\end{table}

Since the UVOT calibration is based on counts measured within a 5 arcsec
aperture, a correction must be made if a different aperture is used. 
The size of the correction depends on the PSF of the source of interest which in turn 
depends on the filter being used. The PSF is also observed to vary slightly
throughout  the orbit as the instrument changes temperature. 
The CalDB contains a set of `average' PSFs for each filter 
which can be used to derive a correction; an `average' magnitude correction has 
been derived and included in Table~\ref{apercortable} for quick reference. The
HEAsoft {\it Swift} software tools {\tt uvotsource} and {\tt uvotapercorr} use these PSFs to
derive corrections for any specified user aperture. However, if this method is to be
used, we caution against using an aperture smaller than 3 arcsec because
the orbital variations in the PSF have a significant influence on photometry
calculated with smaller apertures. 

An alternative method that eliminates the problem of the small orbital variation,
is to derive the aperture correction on an exposure by exposure basis. This
also automatically copes with the PSF filter dependency. Furthermore, during settling
exposures, or very occasionally when there are attitude problems, the pointing
can drift during the exposure, causing the images to be blurred. In these
cases an exposure-specific correction is essential. 
The shape of the UVOT PSF also depends
on the count rate of the source;  bright sources ($\ge 10 {\rm~counts\,s^{-1}}$) have narrower
PSFs than faint ones because of coincidence loss effects. This count rate PSF variation could be
compensated for by deriving separate aperture corrections for sources
with different count rates. In practice both the orbital variation and count rate
dependency changes the aperture correction by
only a few percent, except for very high count rates, in which case a
small aperture is not appropriate.   

We present here a suggested method for obtaining well
calibrated data from UVOT, using a small aperture plus an exposure-specific aperture 
correction, and include an example of this method below. However, we do 
not wish to be overly prescriptive, or to tie the user to any particular 
software tools. The reader should not be afraid to experiment with 
different apertures or background regions for different situations. 

One method of aperture correcting for each exposure is as follows:

\begin{enumerate}
\item Identify a set of 5--15 isolated stars in the exposure.  These
stars should be bright enough that there is signal in the wings of the
PSF, but not so bright that coincidence loss 
is distorting the PSF (i.e. between a few and about 10 counts per second). There should
be no neigbouring sources detectable within at least 10 arcsec to avoid using stars 
that are contaminated by light from other sources. 
\item Perform photometry on each of these stars with both the user-supplied
aperture and the 5 arcsec standard photometric aperture. 
\item Subtract the magnitudes in the user-supplied aperture from the magnitudes in the
5 arcsec aperture to get the aperture corrections.  
\item The mean of the 5--15 aperture corrections thus obtained can then be used as the aperture
correction for that exposure. To estimate the error in the aperture
correction take the RMS of the individual values about the mean value. 
\item  The aperture correction is added to the 
magnitudes of the sources of interest measured in the user-supplied aperture in
the same exposure. 
\end{enumerate}

The error in the aperture correction factor for a single measurement is not
simply the square root sum of the measurements at the chosen and the standard
aperture. That is because the photons in the smaller aperture are counted in
measurements with both apertures. It can be shown that the error in the aperture correction
factor is the measurement error in the largest aperture. It is useful to use
this error for deriving the weighted average from multiple measurements,
especially if the stellar magnitudes of the reference stars vary. 

\subsection{Using an aperture correction}
\label{example}
The stars in the PG1633+099B field have been analysed and calibrated using the 
exposure-specific
method outlined above. The results (shown in
Figure~\ref{figScotts}) can be compared
directly with Stetson's photometry giving confidence that the method works
well. The mean offset between Stetson and UVOT photometry in this data set 
is $<0.01$ mag, with
a standard deviation of 0.06 magnitudes for stars brighter than 17th magnitude,
and 0.18 magnitudes if all the stars are included.

\begin{figure}
\includegraphics[angle=-90,width=80mm]{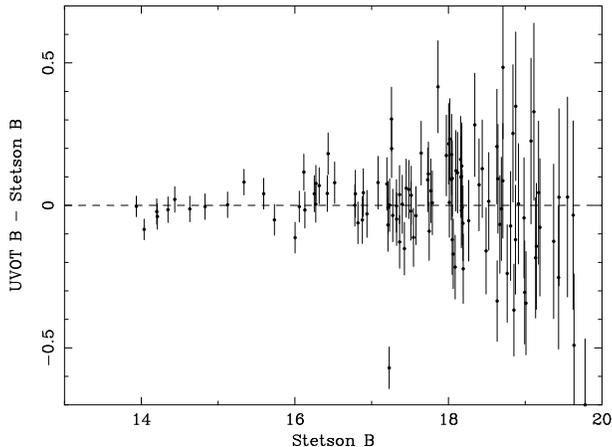}
\caption{A comparison of magnitudes from UVOT data of the PG1633+099B field, 
obtained using the exposure-specific aperture
  correction method outlined in Section~\ref{practical}, with those from 
\citet{b21}. }
\label{figScotts}
\end{figure}

We tested the simple standard PSF approach as used by the {\tt uvotapercorr}
tool in comparison with the exposure-specific method described above. 
The comparison was done for two
v-filter data sets.  We used {\tt
uvotsource} (which calls {\tt uvotapercorr}) to perform photometry on 
several isolated stars in each
frame using both 3 arcsec and 5 arcsec circular apertures.  An
aperture correction to convert from 3 arcsec magnitudes to
5 arcsec magnitudes was computed for each frame and compared to the
aperture correction derived from the v filter PSF given in the CalDB.  We found that the
aperture correction derived from the PSF for these two fields underestimated 
the brightness of the point sources by 0.020 mag in the 00055751001
exposures and 0.019 mag in the 00276321001 exposures. These examples confirm
the fact that the aperture corrections derived from the PSFs are
reliable to a precision of a few hundredths of a magnitude, but that
aperture corrections specific to each exposure need to be computed if
high-precision photometry is desired. Similar offsets are found for different
combinations of filter and aperture size.

\section{Discussion}
\label{discussion}

The approach we have taken in this work 
is superior in a number of ways to that which was used to 
produce the initial in-orbit UVOT photometric calibration \citep{b5}. 
A full 
description of the earlier calibration is beyond the scope of this paper, but
we note here three particularly significant improvements: 

\begin{enumerate}
\item The shape of the instrument response curve used in this work, which is based on
measurements of the individual UVOT optical elements, is a much closer
representation of the true instrument response than that used for the original
calibration, which, although it was a direct measurement and consistent with the
current instrument response, was not of high-precision, and was 
sparsely sampled in wavelength.
\item The effective areas and zero points are determined for the same aperture
for all filters (5 arcsec radius), that is optimised to minimize the 
enclosed-energy dependence on coincidence loss and hence count rate. In our 
initial calibration the effective areas and zero points were based on different 
(and less-optimum) apertures for the UV and optical filters.
\item
The zero points for optical and UV filters are generated using a single 
procedure in the new calibration, whereas in the initial calibration the
optical zero points were obtained such that the mean difference was zero 
between UVOT and Johnson magnitudes of a group of standard
stars: in effect calibrating the UVOT as though its optical filters had 
a response identical to the Johnson system. Defining a UVOT-based magnitude system, and then facilitating conversion to the Johnson system 
using colour transformations gives an inherently more accurate, stable, and 
understandable calibration. 

For comparison with previously-released colour transformations the u-v to U-V fit
for Pickles stars (as shown in the bottom panel of 
Figure~\ref{FigColor}) has been plotted again in Figure~\ref{FigOldCols}, but this time we have added the
initial in-orbit calibration from CalDB 20050805 (dashed line) and also the colour correction provided in
\citet{LW2006} (dash-dot line). \citet{LW2006} suggest no colour correction for the v and b
filters, which would therefore be represented by the dashed horizontal line in
the middle panel of Figure~\ref{FigColor}. The GRB models and fit are not
included in this comparison plot because this is the first time colour transformations have been
provided for GRBs.
\end{enumerate}

\begin{figure}
\includegraphics[angle=0,width=84mm]{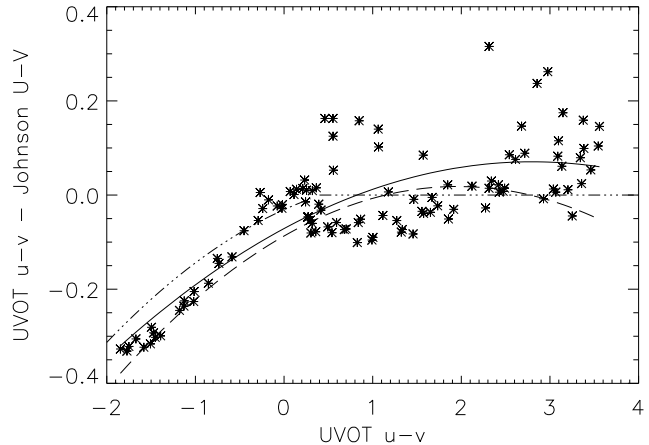}
\caption{UVOT u-v compared with Johnson U-V for Pickles stars (stars) as
  plotted in Figure~\ref{FigColor}.   
The solid line represents the second order polynomial
  fit to the Pickles stars given in Section~\ref{colour trans}, the dashed
  line prepresents the fit given in the initial calibration (CalDB released
  20050805). 
The transformation given by \citet{LW2006} is shown by the dash-dot line.}
\label{FigOldCols}
\end{figure}

For a full list of changes to the calibration approach, procedures and 
observations, the reader is referred to the calibration documentation on the
{\em Swift} website. The CalDB documentation and files containing the
calibrations presented in this paper are
available from the HEASARC CalDB at
http://swift.gsfc.nasa.gov/docs/heasarc/caldb/swift/ in version 20070627 of the
Swift/UVOTA CalDB. In the event that the reader wishes to compare new results with data processed
with previous versions of the pipeline or calibration products, they are
advised to reprocess the older data. The pipeline has been revised to take care
of exposure time problems and to flag any remaining problems.

\subsection{Future photometric calibration}
\label{future}

Although we consider the photometric calibration of UVOT as presented here to
be in good shape, there are a number of areas in which we are working to 
further improve and refine the calibration. For example, we plan to measure the
sensitivity variation over the detector with a higher spatial resolution than
was possible using the observations described in Section~\ref{uniform}. 
For this purpose, we have made a sequence of 55 observations of a dense star 
field (more than 300 stars) at a range of offsets and orientations, so that 
each individual star is observed at least 30 times, each at different 
locations on the detector. The high density of data
across the detector and large number of stars will allow us to model the
detector response spatially to minimize photometric error. The goal is to use
these observations to construct a flat-field calibration product 
which will enable us to achieve a photometric response that is uniform to 1--2 
per cent over the entire detector area. 

It would be desirable to increase the number of
standard stars and the colour range on which the UV and white filters are 
calibrated. The main limitation is the lack of UV spectrophotometric standard 
stars which are sufficiently
faint for UVOT. This may be overcome in two ways: by using smaller hardware
windows we can bring some brighter standards within the range of calibrated
coincidence loss, and also we would like to make use of white dwarfs from the 
Sloan Digital Sky Survey~\citep{i2} for which there are accurate models of the 
UV flux. 

We also plan to characterise the point spread
function out to a large radius to enable, for example, reliable surface 
photometry. 

The detector response is expected to decline gradually 
with total radiation dose. From experience with XMM-OM we expect this to be at
a level of 1--2 percent per year. Although this decline is not yet large enough to be
measured, continued observations of standard stars will allow us to measure and
calibrate the changing photometric response of UVOT.

There is also more work to be done on various aspects of practical photometry,
for instance, by developing an optimal sky-fitting algorithm to remove stars in 
the background region, given that the background is usually in the Poisson 
statistics regime rather than the normally-assumed Gaussian. Also, the
parameterization of the PSF with count rate and with orbital variations will lead to the
development of more accurate aperture corrections.

\subsection{Summary}
\label{summary}

In this paper we have presented the in-orbit photometric calibration of 
the UVOT and defined the UVOT photometric system. We have discussed
factors which affect the accuracy of the photometry.
Any observational science depends critically on the calibration of the
instruments; the UVOT calibration is itself based on trusted sources
and data processed with the most up-to-date software. The comparison with other
calibrated data sets gives us confidence that we have a robust, reliable and accurate system.

\section*{Acknowledgments}
We would like to thank Weidong Li, Joshua Bloom and Alexei Filippenko for
testing the calibration and for useful discussions.
{\it Swift} UVOT was designed and built in collaboration between MSSL, PSU, SwRI,
Swales Aerospace and GSFC, and was launched by NASA. We would like to thank all
those involved in the continued operation of UVOT at PSU, MSSL and GSFC, and
those involved in the data processing and the writing of analysis software. 
This work is supported at MSSL by funding from PPARC and at PSU by NASA's
Office of Space Science through grant NAS5-00136. 
Some of the standard spectra used in
this paper were obtained from the Multimission Archive at the Space Telescope
Science Institute (MAST). STSci is operated by the Association of Universities
for Research in Astronomy, Inc., under NASA contract NAS5-26555. Support for
MAST for non-HST data is provided by the NASA Office of Space Science via grant
NAG5-7584 and by other grants and contracts.
We acknowledge the use of public data from the {\it Swift} data archive.

\end{document}